\documentclass[sigconf]{acmart}

\usepackage[linesnumbered,ruled,vlined]{algorithm2e}
\usepackage{amsthm,graphics,float,subfigure}
\usepackage{bm}
\usepackage{url}
\usepackage{multirow}
\usepackage{verbatim}
\usepackage{color}
\newcommand{\stitle}[1]{\vspace{1ex} \noindent{\bf #1}}
\long\def\comment#1{}
\newcommand{\kw}[1]{{\ensuremath{\mathsf{#1}}}\xspace}

\newcommand{\bc}{\kw{BCList\text{++} }}

\newcommand{\pivot}{\kw{PIVOTER}}
\newcommand{\edgepivot}{\kw{EPivoter}}
\newcommand{\npivot}{\kw{NPivoter}}
\newcommand{\npc}{\kw{NPC}}
\newcommand{\cnt}{\kw{count}}
\newcommand{\lcnt}{\kw{localcount}}
\newcommand{\rcnt}{\kw{rangecount}}

\newtheorem{proof sketch}{PROOF SKETCH}

\begin{document}

\title{Fast Biclique Counting on Bipartite Graphs: A Node Pivot-based Approach}

\author{Xiaowei Ye, Rong-Hua Li, Longlong Lin, Shaojie Qiao, Guoren Wang}

\begin{abstract}
%$(p,q)$-biclique $(X, Y)$ is a complete subgraph in bipartite graph with $|X|=p, |Y|=q$. The problem of $(p,q)$-biclique counting and enumeration is a fundamental task for many applications. However, the existing algorithms for this problem are often very costly for real-world networks. We propose a novel idea of node-pivot, with which we develop a novel algorithm \npivot, a general biclique counting framework. The previous works are all special implementations of the framework. We also provide a novel implementation of \npivot which has better worst-case time complexity. Despite only counting $(p,q)$-biclique, \npivot also applies for local counting (compute the count of $(p,q)$-biclique for each node) and range counting (compute the count of bicliques with size in a range simultaneously). Extensive experiments on 12 real-world graphs demonstrate that \npivot  an outperform the state-of-the-art algorithm by up to $2$  orders of magnitude.

Counting the number of $(p, q)$-bicliques (complete bipartite subgraphs) in a bipartite graph is a fundamental problem which plays a crucial role in numerous bipartite graph analysis applications. However, existing algorithms for counting $(p, q)$-bicliques often face significant computational challenges, particularly on large real-world networks. In this paper, we propose a general  biclique counting framework, called \npivot, based on a novel concept of node-pivot. We show that previous methods  can be viewed as specific implementations of this general framework. More importantly, we propose a novel implementation of \npivot based on a carefully-designed minimum non-neighbor candidate partition strategy.  We prove that our new implementation of \npivot has lower worst-case time complexity than the state-of-the-art methods.  Beyond basic biclique counting, a nice feature of \npivot is that it also supports local counting (computing bicliques per node) and range counting (simultaneously counting bicliques within a size range). Extensive experiments on 12 real-world large datasets demonstrate that our proposed \npivot substantially  outperforms state-of-the-art algorithms by up to two orders of magnitude.
\end{abstract}

\maketitle

\section{introduction}

A bipartite graph is a special type of graph where the node set can be divided into two disjoint sets, typically denoted as \( U \) and \( V \), such that every edge in the graph connects a node from \( U \) to a node from \( V \), with no edges within \( U \) or \( V \). If \( X \subseteq U \) and \( Y \subseteq V \), the pair \((X, Y)\) forms a \((p, q)\)-biclique if \( |X| = p \), \( |Y| = q \), and every node in \( X \) is connected to every node in \( Y \). 

The count of \((p, q)\)-biclique is a key structural measure in bipartite graphs. In graph database query optimization, this count helps predict query processing costs by estimating the result set size in join operations, improving query planning and execution efficiency \cite{DBLP:journals/pvldb/MhedhbiS19,DBLP:journals/tods/AbergerLTNOR17}. In graph neural networks (GNNs), counting \((p, q)\)-bicliques for each node captures higher-order graph structural properties, which are  crucial for improving model performance \cite{GNNlocalCount, qian2022ordered,  DBLP:conf/kdd/YanZG0Z24, DBLP:journals/pami/BouritsasFZB23, DBLP:journals/vldb/ZhaoYLZR23}. \((p, q)\)-biclique counts are also essential in various real-life applications, such as social network analysis \cite{19vldbdensegraph,cbk2011Tolerating,kdensWWW2015} and bioinformatics \cite{bkmotifInbioinformatics}, where they help identify dense substructures representing highly connected clusters of users, entities, or data items.

%The sota methods for \((p, q)\)-biclique counting are \bc \cite{BClistYangPZ21} and \edgepivot \cite{edgepivot}. \bc is a listing-based algorithm. The idea of \bc is to enumerate all $p$-node subset of $U$, namely $L$, which have $R$ common neighbors in $V$. The count of the \((p, q)\)-biclique is $|R| \choose q$. \bc provides pruning techniques to avoid unnecessary search. \edgepivot observes that a \((p, q)\)-biclique is usually a subset of a larger biclique. \edgepivot enumerates edges to build a tree data structure, where Each tree vertex is an edge.   Each from root to leaf tree path combines to a large biclique. This search tree makes sure each  \((p, q)\)-biclique is in exactly one large biclique. Therefore, \edgepivot can enumerate large bicliques to count \((p, q)\)-biclique. Both \bc and \edgepivot split the whole bipartite graph into smaller subgraphs, and then count \((p, q)\)-biclique in these subgraphs, but they use different graph-split strategies. Since \bc enumerates nodes, \bc split the graph according to the neighbors of nodes. Since \edgepivot enumerates edges, \edgepivot split the graph according to the neighbors of edges.

The state-of-the-art (SOTA) methods for \((p, q)\)-biclique counting are \bc \cite{BClistYangPZ21} and \edgepivot \cite{edgepivot}. \bc is a listing-based algorithm that works by enumerating all \(p\)-node subsets of \(U\) (denoted as \(L\)) and finding their common neighbors in \(V\), represented as \(R\). The count of \((p, q)\)-bicliques is then calculated as \( {|R| \choose q} \), where the pruning techniques in \bc help avoid unnecessary searches. \edgepivot, on the other hand, observes that \((p, q)\)-bicliques are often subsets of larger bicliques. It uses edge enumeration to build a tree-like data structure, where each node in the tree represents an edge, and each root-to-leaf path corresponds to a large biclique. This search tree ensures that each \((p, q)\)-biclique appears in exactly one large biclique, allowing \edgepivot to count \((p, q)\)-bicliques by enumerating the larger ones. Both \bc and \edgepivot split the bipartite graph into smaller subgraphs before counting the bicliques, but they employ different graph-split strategies. \bc, focusing on node enumeration, splits the graph based on the neighbors of nodes (namely \textbf{node-split}), whereas \edgepivot, through edge enumeration, splits the graph based on edge neighbors (namely \textbf{edge-split}). This difference in strategy leads to varying efficiencies in counting \((p, q)\)-bicliques depending on the graph structure and parameter values of \(p\) and \(q\).

%While effective, existing algorithms for $(p, q)$-biclique counting still face some limitations. \bc performs well for small values of $p$ and $q$, but its performance degrades rapidly as $p$ and $q$ increase. \edgepivot, while suitable for larger $p$ and $q$ values, can handle range counting (counting bicliques with size in a range simultaneously), but it suffers from high constant factors that reduce its efficiency for smaller values of $p$ and $q$. The high constant factor comes from the enumeration of edges, which leads to more operation of set intersection. Moreover, both \bc and \edgepivot are hindered by high time complexity and rely on a single graph-split strategy, which limits their flexibility and scalability. Moreover, many applications need to compute the count of  $(p, q)$-biclique for each node or edge (local counting), as the property of node or edge. unfortunately, Neither \bc nor \edgepivot suits for local counting.

Despite the success, existing algorithms for \((p, q)\)-biclique counting still have several significant limitations. For example, \bc performs well only when \(p\) and \(q\) are small, but its efficiency diminishes sharply as these values increase \cite{BClistYangPZ21, edgepivot}. On the other hand, \edgepivot is more suited for larger \(p\) and \(q\)  and can handle \textbf{range counting}, i.e., counting $(p,q)$-bicliques for different $p$ and $q$ in a range simultaneously. However, \edgepivot suffers from high constant factors in its time complexity, primarily due to its edge-based enumeration approach. This method leads to more frequent set intersection operations, which increases computational overhead and makes it less efficient for smaller values of \(p\) and \(q\) \cite{edgepivot}. Additionally, both \bc and \edgepivot are limited by their high time complexity and reliance on a single graph-split strategy, which restricts their flexibility and scalability across various graph structures. Moreover, many applications, such as graph neural networks and graph clustering, require \textbf{local counting}, which involves computing the count of \((p, q)\)-bicliques for individual nodes as an important property. Unfortunately, neither \bc nor \edgepivot is well-suited for efficient local counting, further limiting their applicability in real-world scenarios.

%In this paper, we propose \npivot, a general framework for biclique counting. The \npivot framework is general in two ways. First, the \npivot framework provides a general view of biclique counting that both \bc and \edgepivot are special implementations of this framework.  Second, \npivot not only supports basic $(p, q)$-biclique counting but also extends to local counting and range counting.

To overcome the above limitations, we propose \npivot, a versatile and general framework for biclique counting, which provides a theoretically and practically efficient algorithm. The \npivot framework is general in two ways. First, it offers a unified perspective on biclique counting, where both \bc and \edgepivot can be regarded as specific implementations within this framework. This not only establishes \npivot as a significant advancement over existing methods  but also highlights its flexibility in addressing a wider range of cases. Second, \npivot extends beyond traditional \((p, q)\)-biclique counting to support both local counting and range counting. These extensions make \npivot applicable to numerous real-world applications while ensuring greater efficiency and adaptability.

The core idea of the \npivot framework is the concept of node-pivot. A set of nodes are called node-pivots if they connect to all the nodes on the opposite side. In \npivot, these node-pivots simplify the enumeration process  by eliminating the need for further exploration and building a unique representation for all bicliques. This unique representation allows \npivot to efficiently count bicliques in a combinatoric way. On top of that, we develop  an advanced  implementation of the \npivot framework by the candidate set partition strategy called the \textit{minimum non-neighbor candidate partition}. This candidate partition makes the search tree of the \npivot as small as possible. With this candidate partition strategy, we prove that our implementation achieves lower worst-case time complexity than the previous SOTA algorithms. For example, the time complexity of \bc depends on the result size, whereas \edgepivot has a time complexity of \(O(|E|3^{|E_{max}|/3})\), which is independent of the result size. In contrast, our new implementation of \npivot has a result-size-free time complexity of \(O(|E|2^{n'/2})\), where \(2^{1/2} < 3^{1/3}\) and \(n' < |E_{max}|\), making it more efficient in the worst case. Furthermore, we enhance the efficiency of \npivot with a carefully-designed \textit{cost estimator}, which dynamically selects the node-split and edge-split strategies. This hybrid approach adapts to the specific structure of the graph, taking advantage of both node-split and edge-split to reduce computational overhead and improve scalability. In a nutshell, our main contributions are briefly summarized as follows:

%Extensive experimental results validate the superior performance and scalability of \npivot across diverse real-world datasets. 

\stitle{A General Counting Framework:}
We propose \npivot, a versatile and general framework for $(p, q)$-biclique counting, built upon the novel node-pivot technique. This framework not only unifies existing algorithms like \bc and \edgepivot as special cases but also extends beyond them by supporting local and range biclique counting.
	
\stitle{A Powerful Candidate Partition Strategy: } 
We develop a powerful candidate partitioning technique—minimum non-neighbor candidate partition—which optimizes the search tree and achieves lower worst-case time complexity than SOTA approaches.
		
\stitle{An Adaptive Splitting Technique: }
Our framework integrates a cost estimator that dynamically switches between node-split and edge-split strategies. This adaptive approach tailors the biclique counting process to the graph's structure, reducing computational overhead and improving practical performance.

\stitle{Extensive Experimental Evaluation:}
Through comprehensive experiments on 12 real-world datasets, we demonstrate that the proposed \npivot outperforms SOTA algorithms by up to two orders of magnitude in both basic and extended biclique counting tasks, underscoring the high efficiency and scalability of our solutions.

\section{Preliminaries} \label{sec:prem}
%Let $G=(U,V,E)$ be a bipartite graph, where $U$ and $V$ are two disjoint set of nodes and $E\subseteq U\times V$ denotes the set of edges. For each node $u$, its neighbor set in $V$ is defined as $N(u,V)=\{v|e(u,v)\in E, v\in V\}$. For a vertex set $S$, the set of the common neighbors of $S$ in $V$ is defined as $N(S,V)$, i.e., $N(S,V)=\cap_{u\in S}N(u,V)$. Let $d(u,V)$ be the degree of $u$ in $V$, i.e. $d(u,V)=|N(u,V)|$. If the context is clear, $N(u,V)$, $N(S,V)$ and $d(u, V)$ are abbreviated as $N(u)$, $N(S)$ and $d(u)$, respectively. The 2-hop neighbors of a node $u\in U$ is $N(N(u, V), U)$. The 2-hop neighbors of a node $v\in V$ is $N(N(v, U), V)$.  A rank $R$ of a set of nodes $S$ is a vector that maps each node of $S$ into a number in $[1, |S|]$.

Let \( G = (U, V, E) \) be a bipartite graph, where \( U \) and \( V \) are disjoint sets of nodes, and \( E \subseteq U \times V \) is the set of edges. For each node \( u \in U \), its neighbor set in \( V \) is \( N(u, V) = \{v \in V \mid (u, v) \in E\} \). For a set of vertices \( S \subseteq U \), the common neighbors of \( S \) in \( V \) are defined as \( N(S, V) = \bigcap_{u \in S} N(u, V) \). The degree of a node \( u \) in \( V \), denoted by \( d(u, V) \), is the size of its neighbor set: \( d(u, V) = |N(u, V)| \). When the context is clear, we abbreviate \( N(u, V) \), \( N(S, V) \), and \( d(u, V) \) as \( N(u) \), \( N(S) \), and \( d(u) \), respectively. The 2-hop neighbors of a node \( u \in U \) are given by \( \cup_{v\in N(u,V)}{N(v,U)} \). For all the above  concepts, we have symmetrical definitions for vertices in $V$. A rank \( R \) of a set of nodes \( S \) is a vector that assigns each node in \( S \) a distinct number from the range \([1, |S|]\). Below, we give the definition of $(p, q)$-biclique.

\begin{definition}\label{def-biclique}
Given a bipartite graph $G(U,V,E)$, a $(p, q)$-biclique in $G$ is a complete subgraph with a pair of vertex sets $(X,Y)$ where $|X|=p, |Y|=q$.
\end{definition}

%\stitle{Problem definition.} Given a  bipartite graph $G(U,V,E)$ and two integers $p$ and $q$, the problem is to compute the number of $(p,q)$-bicliques in $G$.
%
%The $(p,q)$-biclique counting problem is NP-hard \cite{BClistYangPZ21}, implying that there is no known polynomial-time algorithm that can solve all instances of the problem efficiently.

\stitle{Problem Definition.} Given a bipartite graph \( G(U, V, E) \) and two  integers \( p \) and \( q \), the task is to compute the number of \((p, q)\)-bicliques in \( G \).

The \((p, q)\)-biclique counting problem is NP-hard \cite{BClistYangPZ21}, which implies that no known polynomial-time algorithm can efficiently solve all instances of the problem.

\begin{figure}[t]
	%	\vspace*{0.2cm}
	\begin{center}
		\includegraphics[width=0.5\linewidth]{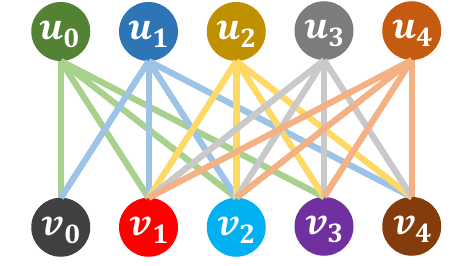}
	\end{center}
	%\vspace*{-0.2cm}
	\caption{An example graph }
	\label{fig:expG}
	%\vspace*{-0.1cm}
\end{figure}
\begin{example}
	Figure~\ref{fig:expG} is a synthetic graph with 10 $(3,3)$-bicliques. For example,  $\left(\{u_0,u_2,u_3,u_4\}_{3}, \{v_1,v_2,v_3\}\right)$ is a $(3,3)$-biclique, where $\{u_0,u_2,u_3,u_4\}_{3}$ indicates choose arbitrary 3 nodes from $\{u_0,u_2,u_3,u_4\}$. The rest of $(3,3)$-bicliques are $(\{u_1,u_2,u_3,u_4\}_{3}, \{v_1,v_2,v_4\})$, \\$(\{u_2,u_3,u_4\}, \{v_1,v_3,v_4\})$, and $(\{u_2,u_3,u_4\}, \{v_2,v_3,v_4\})$.  %$(\{u_0,u_2,u_3\}, \{v_1,v_2,v_3\})$, $(\{u_0,u_2,u_4\}, \{v_1,v_2,v_3\})$, $(\{u_0,u_3,u_4\}, \{v_1,v_2,v_3\})$, $(\{u_1,u_2,u_3\}, \{v_1,v_2,v_4\})$, $(\{u_1,u_2,u_4\}, \{v_1,v_2,v_4\})$, $(\{u_1,u_3,u_4\}, \{v_1,v_2,v_4\})$, $(\{u_2,u_3,u_4\}, \{v_1,v_2,v_3\})$, $(\{u_2,u_3,u_4\}, \{v_1,v_2,v_4\})$, $(\{u_2,u_3,u_4\}, \{v_1,v_3,v_4\})$, $(\{u_2,u_3,u_4\}, \{v_2,v_3,v_4\})$.
\end{example}

\section{Existing Solutions and Their Defects}\label{sec:existingworks}
\subsection{\bc \cite{BClistYangPZ21}} \label{ssec:bc}
%\bc is a listing-based algorithm. Instead of listing each $(p,q)$-biclique one by one, \bc lists the $p$-cliques in a data structure called $2$-hop graph. Given a bipartite graph $G(U,V,E)$, the $2$-hop graph $H(U, E')$ is a directed acyclic graph (DAG) with nodes set $U$ and  every two nodes are linked with an edge if they have over $q$ common neighbors in $V$, i.e. $\forall (u_0, u_1)\in E', |N(\{u_0,u_1\}, V)|\ge q$. For any $(p,q)$-biclique $(X,Y)$, $X$ must be a complete subgraph in $H(U,E')$ because every two nodes in $X$ has $q$ neighbors in $Y\subseteq V$. \bc lists all $p$ nodes complete subgraph in $H(U, E')$ to count $(p,q)$-biclique. For each complete subgraph $X\subseteq U$ that $|X|=p$, the count of $(p,q)$-biclique is $|N(X, V)|\choose q$. Given $G(U,V,E)$ and its 2-hop subgraph $H(U, E')$, the time complexity of \bc is $O(\alpha(H)^{p-2} |E'| d_{max})$ where $\alpha(H)$ is the arboricity of $H$ \cite{CN} and $d_{max}$ is maximum degree in $U(G)$. \bc can also build a $2$-hop graph on $V$ and enumerates the $q$-cliques, and \bc design a estimation model to judge which side is more efficient.

%Instead of listing each \((p,q)\)-biclique individually, \bc
\bc is a listing-based algorithm, which lists the \(p\)-node complete subgraphs in a data structure called the ``2-hop graph''. Specifically, given a bipartite graph \(G(U, V, E)\), the 2-hop graph \(H(U, E')\) is a directed acyclic graph (DAG) with the node set \(U\), where two nodes are connected by an edge if they have at least \(q\) common neighbors in \(V\). Namely, for all \((u_0, u_1) \in E'\), \(|N(\{u_0, u_1\}, V)| \geq q\). Thus, for any \((p, q)\)-biclique \((X, Y)\), the set \(X\) must form a complete subgraph in \(H(U, E')\), as each pair of nodes in \(X\) has at least \(q\) common neighbors in \(Y \subseteq V\). \bc counts the number of \((p, q)\)-bicliques by listing all \(p\)-node complete subgraphs in \(H(U, E')\). For each complete subgraph \(X \subseteq U\) where \(|X| = p\), the number of \((p, q)\)-bicliques is \(\binom{|N(X, V)|}{q}\).

The first search layer of \bc divides the network into smaller subgraphs based on a ranking of nodes and then counts $(p,q)$-bicliques in these subgraphs. Specifically, for each node \(u \in U\), \bc counts the \((p-1, q)\)-bicliques in the subgraph induced by the neighbors and higher-ranked 2-hop neighbors of \(u\). We refer to this method as the \textit{node-split} strategy.

Given the bipartite graph \(G(U, V, E)\) and its 2-hop graph \(H(U, E')\), the time complexity of \bc is \(O(\alpha(H)^{p-2} |E'| d_{\text{max}}+\triangle)\), where \(\alpha(H)\) is the arboricity of \(H\) \cite{CN}, \(d_{\text{max}}\) is the maximum degree in \(U(G)\) and $\triangle$ is the size of result. \bc can also construct a 2-hop graph on \(V\) and enumerate \(q\)-cliques. Additionally, \bc employs an estimation model to determine which side of the graph (either \(U\) or \(V\)) is more efficient for enumeration.

%\bc is efficient for large sparse bipartite subgraph when $p$ or $q$ is small. However, a drawback is that the performance decreases with increasing $p,q$ and graph size. Another limitation is that \bc inputs $p, q$ as parameters and can only count the $(p,q)$-biclique. It needs to run multiple times for each pair of $(p,q)$ when users needs to count the bicliques with size in a range. The third disadvantage is that local counting per node or per edge is not supported, while local counting is required as a node or edge property in some applications, such as graph neural networks \cite{GNNlocalCount}.

\bc is efficient for large, sparse bipartite graphs when \(p\) or \(q\) is small. However, its performance degrades as \(p\), \(q\), or the graph size increases. For instance, on the Twitter dataset ($|U| = 175, 214, |V| = 530, 418, |E| = 1, 890, 661)$, the count of $(5,5)$-biclique reaches $10^{13}$, thereby the listing-based algorithms are doomed to fail. Another limitation is that \bc requires \(p\) and \(q\) as input parameters, meaning that it can only count \((p,q)\)-bicliques for a specific pair of values. To count bicliques across a range of sizes, the algorithm must run multiple times for each pair of \((p,q)\), which is inefficient. A third drawback is that \bc does not support local counting per node, which is essential for certain applications, such as graph neural networks and higher-order graph clustering, where local biclique counts serve as important node properties \cite{GNNlocalCount, qian2022ordered,  DBLP:conf/kdd/YanZG0Z24, DBLP:journals/pami/BouritsasFZB23, DBLP:journals/vldb/ZhaoYLZR23}.

\subsection{\edgepivot \cite{edgepivot}} \label{ssec:edgepivoter}
%Different from \bc, \edgepivot counts the $(p,q)$-biclique in a combinatoric way instead of listing. The straightforward method of combinatoric counting is to enumerate the maximal bicliques first, and then combine the counts of $(p,q)$-biclique in the maximal bicliques. However, since the maximal bicliques are overlapped, a $(p,q)$-biclique may be counted repeatedly.  To solve this problem, \edgepivot encodes a set of large bicliques into a unique representation for each $(p,q)$-biclique by a proposed edge-pivot technique. Specifically, \edgepivot builds a search tree that each tree vertex is an edge and each path from root to leaf is a large biclique. Each of such a large biclique encodes a set of $(p,q)$-bicliques and each $(p,q)$-biclique is encoded by only one large biclique. With the unique representation for each $(p,q)$-biclique, \edgepivot can count $(p, q)$-biclique in a combinatoric way instead of listing one by one.    Given $G(U,V,E)$, the time complexity of \edgepivot is $O(|E|3^{|E_{max}|/3})$ where $|E_{max}| = \max_{(u,v)\in E}{|E\cap (N(u)\times N(v))|}$.

Unlike \bc, which counts \((p,q)\)-bicliques by listing, \edgepivot counts them using a combinatorial approach. A straightforward combinatorial method would first enumerate maximal bicliques and then count the \((p,q)\)-bicliques within them. However, due to significantly overlapping maximal bicliques, a \((p,q)\)-biclique might be counted multiple times. To address this, \edgepivot introduces an edge-pivot technique that encodes large bicliques into a unique representation for each \((p,q)\)-biclique. Specifically, \edgepivot constructs a search tree where each tree vertex represents an edge, and each path from the root to a leaf corresponds to a large biclique. Each large biclique encodes a set of \((p,q)\)-bicliques, and each \((p,q)\)-biclique is uniquely represented by one large biclique. With this unique encoding, \edgepivot can count \((p,q)\)-bicliques combinatorially without listing them one by one.

As discussed above, \bc counts the \((p-1, q)\)-bicliques with the \textit{node-split} strategy. Differently, for each edge \((u, v)\), \edgepivot counts the \((p-1, q-1)\)-bicliques in the subgraphs induced by the higher-ranked neighbors of both \(u\) and \(v\). This method is termed the \textit{edge-split} strategy. Deciding which strategy is more efficient is complex and depends on both the algorithm and the graph structure. Different algorithms and graph structures are better suited to different strategies, and even different parts of a single graph may benefit from various approaches. In this paper, we will propose a novel cost estimator to leverage the advantages of both the node-split and edge-split strategies (Section~\ref{ssec:cost_estimator}).

Given the bipartite graph \(G(U,V,E)\), the time complexity of \edgepivot is \(O(|E| 3^{|E_{\text{max}}|/3})\), where \(|E_{\text{max}}|\)=\(\max_{(u,v)\in E} |E \cap (N(u) \times N(v))|\). Unlike \bc, \edgepivot has a time complexity free from the result size (the $(p, q)$-biclique counts). \edgepivot is more efficient than \bc when \(p\) and \(q\) are relatively large (e.g., $p \ge 6, q\ge 6$), and can count bicliques across a range of sizes simultaneously, thanks for the edge-pivot based  unique encoding \cite{edgepivot}. However, \edgepivot is less efficient when \(p\) and \(q\) are very small (e.g., $p <6, q<6$). The key difference is in the enumeration process: \bc enumerates nodes from one side (i.e., selecting a node from a fixed side during backtracking), while \edgepivot enumerates edges (each search tree vertex represents an edge), selecting two nodes—one from each side—during backtracking. This approach leads to higher constant factors  in its time complexity  for \edgepivot, as edge enumeration requires more set intersection operations, which are computationally intensive. As a result, \edgepivot is not as efficient when \(p\) and \(q\) are very small. Moreover, the time complexity of \edgepivot can be high, as it depends on the parameter \(|E_{\text{max}}|\), which can significantly affect performance.

\section{novel node-pivot based framework}
%As discussed above, \bc enumerates nodes in only one side, which is a simple and slight algorithm. \edgepivot builds a unique representation for all $(p,q)$-bicliques, which is more efficient for large $p$ and $q$. However, \edgepivot suffers from high time complexity, whose exponential part depends on $|E_{max}|$. And \edgepivot also has high constant factors due to enumerating edges. A question is that can we design an  algorithm which (1) keeps simple and slight, (2) builds a unique representation for all $(p,q)$-bicliques, (3) can count the bicliques with size in a range simultaneously and locally count $(p,q)$-bicliques, and (4) has better worst case time complexity? 

As previously discussed, the \bc algorithm enumerates nodes on only one side, which constitutes a straightforward and lightweight approach. On the other hand, \edgepivot constructs a unique representation for all $(p,q)$-bicliques, proving more efficient for relatively larger $p$ and $q$. However, \edgepivot is hindered by a high time complexity, with an exponential component that is contingent upon $|E_{max}|$. On top of that, \edgepivot incurs high constant factors due to its edge enumeration process. A pertinent question arises: is it possible to devise an algorithm that (1) maintains simplicity and lightweightness, (2) is efficient for counting $(p,q)$-biclique whether $p$ and $q$ is large or small, (3) facilitates both local counting and range counting, and (4) exhibits an improved worst-case time complexity?

%\npivot efficiently manages biclique structures.

To achieve these goals, we propose a novel node-pivot based framework \npivot.  Firstly, \npivot employs the proposed powerful node-pivot technique to build a unique representation for all $(p,q)$-bicliques, while the previous work \edgepivot utilizes edge-pivot. Our node-pivot differs significantly from edge-pivot. Our node--pivot is a set of nodes which connect to all the opposite nodes, while edge-pivot is an edge with the maximum number of neighbors. Secondly, since \npivot can construct a unique representation for all $(p,q)$-bicliques, we can use a combinatorial counting approach to significantly reduce redundant calculations. Thirdly, \npivot is a general framework for counting bicliques. We show that both the previous methods \bc and \edgepivot are special implementations of our framework. Further, we propose a specific candidate partition strategy to boost the efficiency and achieve a lower worst-case time complexity (details in Section~\ref{ssec:minimum}). At last, in Section~\ref{ssec:cost_estimator}, we develop a novel cost estimator to judge whether node-split or edge-split is more efficient under different graph structures. 

\subsection{The Proposed \npivot Framework}\label{ssec:npivoter}

\begin{algorithm}[t]
	\caption{ \npivot : Node-pivot based framework}
	\label{alg:npivot}
	\small
	%\scriptsize
	\KwIn{The graph $G(U, V, E)$,  two integers $p$ and $q$}
	\KwOut{The count of $(p,q)$-biclique in $G$}
	\SetKwProg{Fn}{Procedure}{}{}

	Denote $R$ by the rank of nodes in the sorted node set of $U\cup V$\;
	$cnt\gets 0$\;
	\ForEach{$u\in U$}{
		\If(\textit{/*edge split*/}){\textit{/* the cost estimator*/}}{
			\ForEach{$v\in N(u)$}{
				$C_U\gets \{u' | u'\in N(v), R(u')>R(u)  \}$\;
				$C_V\gets \{v' | v'\in N(u), R(v')>R(v)  \}$\;
				
				\npc($C_U, C_V, \emptyset, \emptyset, \{u\}, \{v\}$)\;
			}
		}
		\Else(\textit{/*node split*/}){
			$C_U\gets \cup_{v\in N(u)}\{u' | u'\in N(v), R(u')>R(u)  \}$\;
			\npc($C_U, N(u), \emptyset, \emptyset, \{u\}, \emptyset$)\;
		}
		
	}
	\Return{$cnt$\;}
	
\Fn{\npc($C_U, C_V, P_U, P_V, H_U, H_V$)}{
	\If{$(C_U\times C_V) \cap E = \emptyset$ or $|H_U|=p$ or $|H_V|=q$}  {
		{$\cnt(|C_U|, |C_V|, |P_U|, |P_V|, |H_U|, |H_V|)$};
		\Return\;
	}
	
	Move  nodes that connect to all nodes in $C_V$ from $C_U$ to $P_U$\;
	Move  nodes that connect to all nodes in $C_U$ from $C_V$ to $P_V$\;
	
	Choose $L_U, L_V$ from $C_U, C_V$\;
	\For(\textit{/*enumerate $L_U$*/}){$u\in L_U$}{
		$C_U\gets C_U\setminus u$\;
		\npc($ C_U, C_V\cap N(u), P_U, P_V, H_U\cup \{u\}, H_V$)\;
	}
	\textit{/*enumerate $L_V$ like lines~19-21*/}\;

	\npc($C_U, C_V, P_U, P_V, H_U, H_V$)
}

\Fn{$\cnt(c_0,c_1,p_0,p_1,h_0,h_1)$} {
	{\textbf{if}} {$c_0=0$ or $c_1==0$} {\textbf{then}} {
		 {$cnt\gets cnt+{p_0+c_0\choose p-h_0} {p_1+c_1\choose q-h_1},\Return$}\;
	}

	$cnt\gets cnt+ {p_0+c_0\choose p-h_0} {p_1\choose q-h_1} + {p_0\choose p-h_0} {p_1+c_1\choose q-h_1}-{p_0\choose p-h_0} {p_1\choose q-h_1}$\;
%	\For{$t = 1$ to $\min(c_1, q-h_1)$}{
%		$cnt\gets cnt + {p_0\choose p-h_0} {p_1\choose q-h_1-t} {c_1\choose t} $\;
%	}
	
}
	
\end{algorithm}

 The key idea of the \npivot framework is the following node-pivot.
\begin{definition}[node-pivot]
	For any $C_U\subseteq U, C_V\subseteq V$, let $P_U=\{u| u\in C_U, N(u, C_V)=C_V\}$ and $P_V=\{v| v\in C_V, N(v, C_U)=C_U\}$. We refer to the nodes in $P_U$ and $P_V$ as node-pivots.
\end{definition}
%Let $C'_U=C_U \setminus P_U$ and $C'_V=C_V \setminus P_V$.  Since node-pivots connect to all nodes in the opposite set, any biclique extended by a node-pivot remains a biclique. An $(x,y)$-biclique in $(C'_U , C'_V )$ becomes an $(x+1,y)$-biclique if extended by a node in $P_U$, and an $(x,y+1)$-biclique if extended by a node in $P_V$. Moreover, an $(x,y)$-biclique in $(C'_U , C'_V )$ becomes an $(x+a,y+b)$-biclique if extended by $a$ nodes in $P_U$ and $b$ nodes in $P_V$. This property allows the \npivot framework to shrink the candidate sets from $C_U$ and $C_V$ to $C'_U $ and $C'_V$.  If there are $c$ $(x,y)$-bicliques in $(C'_U, C'_V)$, we can derive that there are $c{|P_U|\choose p-x}{|P_V|\choose q-y}$ $(p,q)$-bicliques in $(C_U,C_V)$. To compute the count of bicliques in $(C'_U , C'_V)$, \npivot splits $(C'_U , C'_V)$ into several subsets. Each subset, namely $(C''_U, C''_V)$, has its own node-pivots and can recursively builds down. Recursively, using node-pivots, \npivot can create a unique representation for all $(p,q)$-bicliques and count them combinatorially.

\stitle{An In-depth Observation.} Given that \(C'_U = C_U \setminus P_U\) and \(C'_V = C_V \setminus P_V\), any biclique extended by a node-pivot preserves its biclique structure since node-pivots are connected to all nodes in the opposite set. For instance, an \((x, y)\)-biclique in \((C'_U, C'_V)\) becomes an \((x+1, y)\)-biclique if extended by a node in \(P_U\) and an \((x, y+1)\)-biclique if extended by a node in \(P_V\). Furthermore, an \((x, y)\)-biclique can evolve into an \((x+a, y+b)\)-biclique by extending it with \(a\) nodes from \(P_U\) and \(b\) nodes from \(P_V\). This important characteristic of node-pivots enables the \npivot framework to significantly reduce the size of the candidate sets by shrinking \(C_U\) and \(C_V\) to \(C'_U\) and \(C'_V\), respectively. If there are \(c\) \((x, y)\)-bicliques in \((C'_U, C'_V)\), the total number of \((p, q)\)-bicliques in \((C_U, C_V)\) can be computed by \(c \binom{|P_U|}{p - x} \binom{|P_V|}{q - y}\), in a combinatorial way. To calculate the bicliques in \((C'_U, C'_V)\), \npivot divides \((C'_U, C'_V)\) into multiple smaller subsets. Each of these small subsets has its own node-pivots, and the \npivot framework recursively processes them to build down the problem. Through this recursive approach, \npivot efficiently represents and counts all \((p, q)\)-bicliques combinatorially.

The process of dividing \((C'_U, C'_V)\) into smaller subsets, referred to as the candidate partition problem, has a significant impact on the performance of the \npivot framework. Given four node sets \(L_U \subseteq C_U \subseteq U\) and \(L_V \subseteq C_V \subseteq V\), the bicliques in the bipartite graph \(G(C_U, C_V)\) can be classified into two main categories: those that include nodes from \(L_U\) and \(L_V\), and those that do not. The \npivot framework enumerates \(L_U\) and \(L_V\) to break the candidate sets into multiple subsets. Each subset counts the bicliques that include nodes from \(L_U\) and \(L_V\). The final subset, \(C_U \setminus L_U\) and \(C_V \setminus L_V\), is used to count the bicliques that do not include nodes from either \(L_U\) or \(L_V\). We propose a specific candidate partition strategy that ensures \npivot operates within a worst-case time complexity (details in Section~\ref{ssec:cost_estimator}).

\stitle{Remark.} We will show later that both \bc and \edgepivot can be viewed as specific implementations of the broader \npivot framework, each with distinct strategies for selecting node-pivots.  \bc processes the graph without leveraging node-pivot based pruning. On the other hand, \edgepivot prunes an edge \((u,v)\) where \(u \in P_U\) and \(v \in P_V\). In contrast, the \npivot framework prunes all node-pivots, which enables it to prune a greater number of candidate sets compared to \bc and \edgepivot.  As demonstrated in the experiments discussed in Section~\ref{sec:exp}, over 85\% of \((10,10)\)-bicliques across the tested graphs are counted using this combinatorial method, showcasing the effectiveness of the \npivot approach.

\stitle{Implementation Details.}
Algorithms~\ref{alg:npivot} shows the pseudo-code of \npivot. \npivot inputs $G(U,V,E), p, q$ and outputs the count of $(p,q)$-biclique in $G$. Let $R$ be the rank of nodes (line~1).  \npivot splits the bipartite graph into smaller subgraphs and counts the $(p,q)$-bicliques within them. There are two splitting strategies. The first is the edge-split, where an edge $(u, v)$ is chosen, and the algorithm counts $(p-1, q-1)$-bicliques in the subgraph induced by nodes of higher rank (lines 5-8 of Algorithm~\ref{alg:npivot}). The second is the node-split, where a node $u$ is chosen, and the algorithm counts $(p-1, q)$-bicliques in the subgraph induced by $N(u)$ and higher-ranked 2-hop nodes (lines 10-11 of Algorithm~\ref{alg:npivot}). To determine which strategy is more efficient (line~4), we provide a cost estimation technique, detailed in Section~\ref{ssec:cost_estimator}.

The \npc procedure is central to the \npivot framework (line~13). It takes six parameters: $C_U$, $C_V$, $P_U$, $P_V$, $H_U$, and $H_V$. Here, $C_U$ and $C_V$ represent the candidate sets, where the nodes are common neighbors of $P_V \cup H_V$ and $P_U \cup H_U$, respectively. The sets $P_U$, $P_V$, $H_U$, and $H_V$ form a partial biclique, where $(P_U \cup H_U, P_V \cup H_V)$ is already a biclique, and $|H_U| < p$ and $|H_V| < q$. The goal of \npc is to count the $(p,q)$-bicliques that contain all of $H_U$ and $H_V$, and partially include nodes from $C_U$, $C_V$, $P_U$, and $P_V$. Importantly, $P_U$ and $P_V$ serve as the node-pivots.
%Since node-pivots connect to all the candidate nodes, each biclique added by the node-pivots are still a biclique. Thus, after the removal of the node-pivots from $C_U$ and $C_V$ to $P_U$ and $P_V$ (lines~16-17), we only need to count the bicliques in the remaining $C_U,C_V$. For example, if a node $u$ is removed from $C_U$ to $P_U$ as a node-pivot and there are $x$ $(p-1,q)$-bicliques in the remaining candidate sets, it is easy to derive that there are $x$ $(p,q)$-bicliques directly. Thus, we only need to compute the count of biclique in the remaining candidate sets (lines~18-23). 
Since node-pivots connect to all candidate nodes, any biclique extended by node-pivots remains a biclique. Therefore, after moving node-pivots from $C_U$ and $C_V$ to $P_U$ and $P_V$ (lines~16-17), the task reduces to counting bicliques in the remaining $C_U$ and $C_V$. For example, if a node $u$ is moved from $C_U$ to $P_U$ as a node-pivot and there are $x$ $(p-1,q)$-bicliques in the remaining candidate sets, there will be $x$ $(p,q)$-bicliques directly. Thus, the algorithm focuses on counting bicliques in the remaining candidate sets (lines~18-23).

%After the removal of node-pivots, \npc computes the remaining candidate sets. Instead of enumerating one by one, \npc splits the candidate sets into $L_U,L_V$ and $C_U\setminus L_U, C_V\setminus L_V$ (line~18). The bicliques in $(C_U,C_V)$ can be classified according to whether contains nodes in $L_U, L_V$. For each node $u$ in $L_U$ and $L_V$, \npc recursively enumerates the bicliques that contains $u$ and does not contain the nodes before $u$ (lines~19-22). Note that the nodes are removed from the candidate sets to avoid repeated counting (line~20).  The nodes that must be included are placed in $H_U$ and $H_V$ (line~21). Let $u_0,u_1$ and $u_2$ be three nodes in line~19. For $u_0$, the recursive call in line~21 counts all the bicliques that must contain $u_0$ and may contain $u_1, u_2$. For $u_1$, the recursive call in line~21 counts all the bicliques that do not contain $u_0$, must contain $u_1$, and may contain $u_2$. 

After removing node-pivots, \npc processes the remaining candidate sets. Instead of enumerating one node at a time, \npc splits the sets into $L_U$, $L_V$ and their complements $C_U \setminus L_U$ and $C_V \setminus L_V$ (line~18). The bicliques in $(C_U, C_V)$ are categorized based on whether they include nodes from $L_U$ and $L_V$. For each node in $L_U$ and $L_V$, \npc recursively counts bicliques that include the node and exclude the nodes preceding it (lines~19-22). The nodes are removed from the candidate sets to avoid double-counting (line~20), and the nodes that must be included are placed in $H_U$ and $H_V$ (line~21). For instance, if $u_0$, $u_1$, and $u_2$ are three nodes considered in line~19, the recursive call in line~21 counts bicliques that must include $u_0$ and may include $u_1$ and $u_2$. The next recursive call counts bicliques that exclude $u_0$, must include $u_1$, and may include $u_2$. Finally, after removing $L_U$ and $L_V$ from $C_U$ and $C_V$ (lines~20 and 22), \npc recursively counts the bicliques in the remaining candidate sets (line~23).

%In line~23, $L_U,L_V$ are already removed from $C_U,C_V$ (line~20 and 22), and \npc count the bicliques in the remaining $C_U,C_V$ recursively.

%Thanks for the node-pivots, the large biclique $(P_U\cup H_U, P_V\cup H_V)$ encodes a unique representation of the set of bicliques $\{(P'_U\cup H_U, P'_V\cup H_V) | P'_U\subseteq P_U, P'_V\subseteq P_V \}$. In other words, $H_U,H_V$ are required to be chosen and $P_U, P_V$ are optional. Thus, we can derive Theorem~\ref{the:combination}.

Thanks to the node-pivots, the large biclique $(P_U \cup H_U, P_V \cup H_V)$ encodes a unique representation of the set of bicliques $\{(P'_U \cup H_U, P'_V \cup H_V) \mid P'_U \subseteq P_U, P'_V \subseteq P_V \}$. In this framework, $H_U$ and $H_V$ are required to be chosen, while $P_U$ and $P_V$ are optional. This leads to the derivation of Theorem~\ref{the:combination}.

\begin{theorem}\label{the:combination}
	There are ${|P_U|\choose p-|H_U|}\times{|P_V|\choose q-|H_V|}$ $(p,q)$-bicliques encoded in a large biclique $(P_U\cup H_U, P_V\cup H_V)$. 
\end{theorem}

%\npc terminates when there is no edge or the number of nodes that must be included reaches the max size (lines~14-15). There are four cases as described in the following, all of which are based on Theorem~\ref{the:combination}.

The \npc procedure terminates when either there is no edge left or the number of required nodes reaches the maximum size (lines 14-15). The following four cases, based on Theorem~\ref{the:combination}, describe the scenarios of the \cnt procedure (lines~14-15 and 24-26).

\begin{itemize}
	\item[Case~1: ] $(C_U \times C_V) \cap E = \emptyset$, with $C_U \neq \emptyset$ and $C_V \neq \emptyset$.  
	Here, $C_U$ represents the common neighbors of $P_V \cup H_V$, so all nodes in $C_U$ can be moved into $P_U$. The $(p,q)$-biclique in $((P_U \cup C_U) \cup H_U, P_V \cup H_V)$ is counted as ${p_0 + c_0 \choose p - h_0} {p_1 \choose q - h_1}$ (line 26). Similarly, $C_V$, being the common neighbors of $P_U \cup H_U$, gives the item ${p_0 \choose p - h_0} {p_1 + c_1 \choose q - h_1}$. However, directly adding $C_V$ to $P_V$ would result in double-counting the pair $(P_U \cup H_U, P_V \cup H_V)$, so we subtract ${p_0 \choose p - h_0} {p_1 \choose q - h_1}$.
	\item[Case~2:] $(C_U \times C_V) \cap E = \emptyset$, with $C_U = \emptyset$.  
	In this case, all nodes in $C_V$ are moved to $P_V$, and the $(p,q)$-biclique in $((P_U \cup C_U) \cup H_U, (P_V \cup C_V) \cup H_V)$ is counted (line 25).
	\item [Case~3:]  $(C_U \times C_V) \cap E = \emptyset$, with $C_V = \emptyset$.  
	This case mirrors Case 2.
	\item [Case~4:] $|H_U| = p$ or $|H_V| = q$. When $|H_U| = p$ or $|H_V| = q$, we have $p - h_0 = 0$ or $q - h_1 = 0$, and the counts on lines 25-26 are correct, as they account for the completed bicliques.
\end{itemize}

\begin{theorem}\label{the:correct}
	\npivot can count $(p,q)$-bicliques correctly.
\end{theorem}
\begin{proof}
	%We prove the theorem by stating the fact that each $(p,q)$-biclique is counted once time. Let the recursive search process of the \npc procedure be a search tree. At each search tree vertex, each biclique is counted in only one child node. If the biclique contains nodes in $L_U\cup L_V$, it will be counted in line~21 or 22. Since the nodes $L_U\cup L_V$ are removed (line~20), there is no repeated counting. If the biclique does not contain nodes in $L_U\cup L_V$, it will be counted in the child node from line~23. Thus, each biclique is in only one path in the search tree and counted once time.
	
To prove Theorem~\ref{the:correct}, it is sufficient to show that each $(p,q)$-biclique is counted exactly once. Consider the recursive search process of the \npc procedure as a search tree. At each search tree vertex, every biclique is counted in exactly one root to leaf path. At each vertex in the search tree,	if a biclique contains nodes from $L_U \cup L_V$, it will be counted in lines 21 or 22. Since the nodes in $L_U \cup L_V$ are removed from the candidate sets (line 20), no repeated counting occurs for those nodes. If a biclique does not contain nodes from $L_U \cup L_V$, it will be counted in the child tree vertex generated from line 23. Thus, each biclique belongs to only one path in the search tree and is counted exactly once. This ensures that the algorithm correctly counts all $(p,q)$-bicliques without duplication.
\end{proof}

\stitle{\bc as a Special Implementation of NPivoter.} %Although \bc inputs the whole graph, the first layer of \bc can be seen as the node-split strategy \cite{BClistYangPZ21}. The enumeration process of \bc is a special implementation of the \npc procedure. \bc considers no node-pivots in line~16-17 of Algorithm~\ref{alg:npivot}. \bc sets $L_U=C_U$ and $L_V=\emptyset$ in line~18. There is also no remaining candidate sets in line~23 of Algorithm~\ref{alg:npivot} for \bc. \bc terminates only when $|H_U|=p$ (line~14 of Algorithm~\ref{alg:npivot}). In the \cnt procedure (line~24), \bc has $c_0=p_0=p_1=h_1=0$, $h_0=p$ and $c_1=|C_V|$.
\bc can be viewed as a specialized implementation of the \npivot framework. While \bc processes the entire graph, its initial layer mirrors the node-split strategy  \cite{BClistYangPZ21}. The enumeration process in \bc is a simplified version of the \npc procedure. Specifically, \bc does not utilize node-pivots, as seen in lines 16-17 of Algorithm~\ref{alg:npivot}. It sets $L_U = C_U$ and $L_V = \emptyset$ in line 18, and there are no remaining candidate sets in line 23 for further enumeration. \bc only terminates when $|H_U| = p$ (line 14). During the \cnt procedure (line 24), \bc simplifies the process with $c_0 = p_0 = p_1 = h_1 = 0$, $h_0 = p$, and $c_1 = |C_V|$.

\stitle{\edgepivot as a Special Implementation of NPivoter.}  %Differently,  \edgepivot use the edge-split strategy (lines~5-8 of Algorithm~\ref{alg:npivot}). The enumeration process of \edgepivot is also a special implementation of the \npc procedure.  \edgepivot searches the edge $(u,v)$ that has the maximum number of neighbors and sets $L_U=C_U\setminus N(v) \setminus \{u\}, L_V=C_V\setminus N(u)\setminus \{v\}$. The edge $(u,v)$ is removed into $P_U$ and $P_V$ as an edge-pivot \cite{edgepivot} at line~23, while our \npivot moves all the node-pivots into $P_U$ and $P_V$. In the \cnt procedure, our \cnt procedure utilize a constant-time inclusion-exclusion principle to avoid repeated counting (line~26), while \edgepivot use a enumeration method with linear time complexity.
\edgepivot follows a different approach by using the edge-split strategy (lines 5-8 of Algorithm~\ref{alg:npivot}). Its enumeration process is also a variant of the \npc procedure. \edgepivot selects the edge $(u,v)$ with the maximum number of neighbors and sets $L_U = C_U \setminus N(v) \setminus \{u\}$ and $L_V = C_V \setminus N(u) \setminus \{v\}$. The edge $(u,v)$ will be moved into $P_U$ and $P_V$ as an edge-pivot in the next recursive search layer (line~23). In the \cnt procedure, \npivot uses a constant-time inclusion-exclusion principle to avoid duplicate-counting (line 26), whereas \edgepivot uses an enumeration method with linear time complexity \cite{edgepivot}.

\begin{algorithm}[t]
	\caption{Local counting}
	\label{alg:local}
	\small
	%\scriptsize
	%	\KwIn{The graph $G(U, V, E)$,  two integers $p$ and $q$}
	%	\KwOut{The count of $(p,q)$-biclique in $G$}
	\SetKwProg{Fn}{Procedure}{}{}
	
	$local_u\gets 0, \forall u\in U\cup V$\;
	
	\Fn{$\lcnt(C_U,C_V,P_U,P_V,H_U,H_V)$} {
		%$c_0\gets|C_U|,c_1\gets|C_V|,p_0\gets|P_U|,p_1\gets|P_V|,h_0\gets|H_U|,h_1\gets|H_V|$\;
  $c_0,c_1,p_0,p_1,h_0,h_1\gets|C_U|,|C_V|,|P_U|,|P_V|,|H_U|,|H_V|$\;
		
		\If{$c_0>0$ and $c_1 > 0$}{
			$\lcnt(\emptyset,\emptyset,C_U\cup P_U,P_V,H_U,H_V)$\;
			$P_V\gets P_V\cup C_V$\;
			\For{$v\in C_V$}{
				$P_V\gets P_V\setminus \{v\}$\;
				$\lcnt(\emptyset, \emptyset, P_U, P_V, H_U, H_V\cup \{v\})$\;
			}	
		}
		\Else{
		$p_0\gets p_0+c_0, p_1\gets p_1+c_1$\;
			%\If{$c_0=0$ and $c_1 = 0$}{
				\textbf{for} {$u\in P_U\cup C_U$} {\textbf{do}} {
					$local_u\gets local_u+{p_0-1\choose p-h_0-1} {p_1\choose q-h_1}$\;
				}
				\textbf{for} {$v\in P_V\cup C_V$} {\textbf{do}} {
					$local_v\gets local_v+{p_0\choose p-h_0} {p_1-1\choose q-h_1-1}$\;
				}
				\textbf{for} {$u\in H_U\cup H_V$} {\textbf{do}} {
					$local_u\gets local_u+{p_0\choose p-h_0} {p_1\choose q-h_1}$\;
				}
		
		}
	}
\end{algorithm}

%\begin{algorithm}[t]
%	\caption{Local counting per edge}
%	\label{alg:local_per_edge}
%	\small
%	%\scriptsize
%	%	\KwIn{The graph $G(U, V, E)$,  two integers $p$ and $q$}
%	%	\KwOut{The count of $(p,q)$-biclique in $G$}
%	\SetKwProg{Fn}{Procedure}{}{}
%	
%	$local_{(u,v)}\gets 0, \forall (u,v)\in E$\;
%	
%	\Fn{$\lcnt(C_U,C_V,P_U,P_V,H_U,H_V)$} {
%		%$c_0\gets|C_U|,c_1\gets|C_V|,p_0\gets|P_U|,p_1\gets|P_V|,h_0\gets|H_U|,h_1\gets|H_V|$\;
%		$c_0,c_1,p_0,p_1,h_0,h_1\gets|C_U|,|C_V|,|P_U|,|P_V|,|H_U|,|H_V|$\;
%		
%		\If{$c_0>0$ and $c_1 > 0$}{
%			$\lcnt(\emptyset,\emptyset,C_U\cup P_U,P_V,H_U,H_V)$\;
%			$P_V\gets P_V\cup C_V$\;
%			\For{$v\in C_V$}{
%				$P_V\gets P_V\setminus \{v\}$\;
%				$\lcnt(\emptyset, \emptyset, P_U, P_V, H_U, H_V\cup \{v\})$\;
%			}	
%		}
%		\Else{
%			$p_0\gets p_0+c_0, p_1\gets p_1+c_1$\;
%			%\If{$c_0=0$ and $c_1 = 0$}{
%			$P_U\gets P_U\cup C_U$; $P_V\gets P_V\cup C_V$\;
%				\textbf{for} {$u \in P_U,v\in P_V$} {\textbf{do}} {
%					$local_{(u,v)}\gets local_{(u,v)}+{p_0-1\choose p-h_0-1} {p_1-1\choose q-h_1-1}$\;
%				}
%				\textbf{for} {$u\in P_U, v\in H_V$} {\textbf{do}} {
%					$local_{(u,v)}ets local_{(u,v)}+{p_0-1\choose p-h_0-1} {p_1\choose q-h_1}$\;
%				}
%			\textbf{for} {$u\in H_U, v\in P_V$} {\textbf{do}} {
%				$local_{(u,v)}\gets local_{(u,v)}+{p_0\choose p-h_0} {p_1-1\choose q-h_1-1}$\;
%			}
%				\textbf{for} {$u\in H_U, v\in \cup H_V$} {\textbf{do}} {
%					$local_{(u,v)}\gets local_{(u,v)}+{p_0\choose p-h_0} {p_1\choose q-h_1}$\;
%				}
%				
%			}
%		}
%	\end{algorithm}

\begin{algorithm}[t]
	\caption{Range counting for $  p\in [p_l,p_u], q\in [q_l,q_u]$}
	\label{alg:range}
	\small
	%\scriptsize
	%	\KwIn{The graph $G(U, V, E)$,  two integers $p$ and $q$}
	%	\KwOut{The count of $(p,q)$-biclique in $G$}
	\SetKwProg{Fn}{Procedure}{}{}
	
	$range_{p,q}\gets 0, \forall p\in [p_l,p_u], q\in [q_l,q_u]$\;
	
	\Fn{$\rcnt(C_U,C_V,P_U,P_V,H_U,H_V)$} {
		$c_0,c_1,p_0,p_1,h_0,h_1\gets|C_U|,|C_V|,|P_U|,|P_V|,|H_U|,|H_V|$\;
		$l_0=\max(h_0, p_l), l_1=\max(h_1, q_l)$\;
		\If{$c_0=0$ or $c_1=0$}{
			\For{$p\in [l_0, \min(p_0+c_0+h_0, p_u)]$}{
				\For{$q\in [l_1, \min(p_1+c_1+h_1, q_u)]$}{ 
					$range_{p,q}\gets range_{p,q} + {p_0+c_0\choose p-h_0} {p_1+c_1\choose q-h_1}$\;
				}
			}
		%	\Return\;
		%	$range_{p,q}\gets range_{p,q} + {p_0+c_0\choose p-h_0} {p_1+c_1\choose q-h_1}, \forall p\in [\max(h_0, p_l), \min(p_0+c_0+h_0, p_u)], q\in [\max(h_1, q_l), \min(p_1+c_1+h_1, q_u)]$\;
		}
	\Else{
%		$\rcnt(\emptyset, \emptyset, C_U\cup P_U , P_V, H_U, H_V)$\;

$range_{p,q}\gets range_{p,q}+ {p_0+c_0\choose p-h_0} {p_1\choose q-h_1}, \forall p\in [l_0, \min(p_0+c_0+h_0, p_u)], q\in [l_1, \min(p_1+h_1, q_u)]$\;

$range_{p,q}\gets range_{p,q}+ {p_0\choose p-h_0} {p_1+c_1\choose q-h_1}, \forall p\in [l_0, \min(p_0+h_0, p_u)], q\in l_1, \min(p_1+c_1+h_1, q_u)]$\;

$range_{p,q}\gets range_{p,q}-{p_0\choose p-h_0} {p_1\choose q-h_1}, \forall p\in [l_0, \min(p_0+h_0, p_u)], q\in [l_1, \min(p_1+h_1, q_u)]$\;

%$range_{p,q}\gets range_{p,q}+ {p_0+c_0\choose p-h_0} {p_1\choose q-h_1}, \forall p\in [\max(h_0, p_l), \min(p_0+c_0+h_0, p_u)], q\in [\max(h_1, q_l), \min(p_1+h_1, q_u)]$\;
%
%$range_{p,q}\gets range_{p,q}+ {p_0\choose p-h_0} {p_1+c_1\choose q-h_1}, \forall p\in [\max(h_0, p_l), \min(p_0+h_0, p_u)], q\in [\max(h_1, q_l), \min(p_1+c_1+h_1, q_u)]$\;
%
%$range_{p,q}\gets range_{p,q}-{p_0\choose p-h_0} {p_1\choose q-h_1}, \forall p\in [\max(h_0, p_l), \min(p_0+h_0, p_u)], q\in [\max(h_1, q_l), \min(p_1+h_1, q_u)]$\;
	}

	}
\end{algorithm}

\stitle{Local Counting.} %We can extend Theorem~\ref{the:combination} for local counting. Each node in $H_U\cup H_V$ has ${|P_U|\choose p-|H_U|}\times{|P_V|\choose q-|H_V|}$ $(p,q)$-bicliques. Each node in $P_U$ has ${|P_U|-1\choose p-|H_U|-1}\times{|P_V|\choose q-|H_V|}$ $(p,q)$-bicliques. Each node in $ P_V$ has ${|P_U|\choose p-|H_U|}\times{|P_V|-1\choose q-|H_V|-1}$ $(p,q)$-bicliques. Replacing the \cnt procedure by a new procedure for local counting, \npivot can be used for local counting. We omit the details of the pseudo-code due to the page limit.
Theorem~\ref{the:combination} can be extended to perform local counting. Each node in $H_U \cup H_V$ is involved in ${|P_U| \choose p - |H_U|} \times {|P_V| \choose q - |H_V|}$ $(p,q)$-bicliques. A node in $P_U$ contributes to ${|P_U| - 1 \choose p - |H_U| - 1} \times {|P_V| \choose q - |H_V|}$ $(p,q)$-bicliques, while a node in $P_V$ contributes to ${|P_U| \choose p - |H_U|} \times {|P_V| - 1 \choose q - |H_V| - 1}$ $(p,q)$-bicliques. By modifying the \cnt procedure to perform local counting, \npivot can efficiently handle this task. Algorithm~\ref{alg:local} is the pseudo-code for local counting. When $c_0>0$ and $c_1>0$ (line~4), \lcnt firstly insert $C_U$ into $P_U$ to count the bicliques in $(C_U\cup P_U\cup H_U, P_V\cup H_V)$, i.e. the bicliques without nodes in $C_V$ (line~5), and then count the bicliques containing nodes in $C_V$ (lines~6-9).  When $c_0=0$ or $c_1=0$, insert the cadidate sets into the node-pivots (line~11), and count the bicliques for each kind of nodes (lines~12-14).  %Due to space constraints, the pseudo-code for local counting is omitted.

\stitle{Range Counting.} %Given a $(P_U\cup H_U, P_V\cup H_V)$, Theorem~\ref{the:combination} holds for all $p\in [|H_U|, |P_U\cup H_U|], q\in [|H_V|, |P_V\cup H_V|]$. The \cnt procedure of \npivot only counts for a pair of $p$ and $q$. Thus, with the range $[p_l,p_u]$ and $[q_l,q_u]$, \npivot can count bicliques with the size in the range simultaneously by running the \cnt procedure in $(p_u-p_l)(q_u-q_l)$ times. 
Given a large biclique $(P_U \cup H_U, P_V \cup H_V)$, Theorem~\ref{the:combination} applies to all $p \in [|H_U|, |P_U \cup H_U|]$ and $q \in [|H_V|, |P_V \cup H_V|]$. The current \cnt procedure in \npivot only counts bicliques for a specific pair of $p$ and $q$. To handle range counting with specified ranges $[p_l, p_u]$ and $[q_l, q_u]$, \npivot can execute the \cnt procedure $(p_u - p_l)(q_u - q_l)$ times, enabling it to count bicliques within the given size range simultaneously. Algorithm~\ref{alg:range}
 is the pseudo-code. $l_0$ and $l_1$ are the minimum number of nodes to contain, which is the lower bound of $p$ and $q$ (line~4). When $c_0=0$ or $c_1=0$,  $\min(p_0+c_0+h_0, p_u)$ and $ \min(p_1+c_1+h_1, q_u)$ are the upper bound of $p$ and $q$ (line~6 and 7). When $c_0>0$ and $c_1>0$, lines~10 counts the bicliques in $((C_U\cup P_U)\cup H_U, P_V\cup H_V)$, line~11  counts the bicliques in $( P_U\cup H_U, (C_V\cup P_V)\cup H_V)$,  and line~12 subtract the double-counting part $( P_U\cup H_U,  P_V\cup H_V)$.
 
%Node pivot的思想
%
%部分集合递归枚举的思想
%
%组合计数的思想
%
%证明正确性
%
%bclist++和epivoter都在这个框架下

\begin{algorithm}[t]
	\caption{Minimum non-neighbor candidate partition}
	\label{alg:minnpivot}
	\small
	%\scriptsize
%	\KwIn{The graph $G(U, V, E)$,  two integers $p$ and $q$}
%	\KwOut{The count of $(p,q)$-biclique in $G$}
	\SetKwProg{Fn}{Procedure}{}{}
\Fn{\npc($C_U, C_V, P_U, P_V, H_U, H_V$)}{
	lines~14-17 of Algorithm~\ref{alg:npivot}\;
	
	$w\gets \arg\min_{u\in C_U\cup C_V}{\left (\min(|C_U\setminus N(u)|,|C_V\setminus N(u)|)\right )}$\;
	%	Let $w$ be the minimum non-neighbor node-pivot\;
	$L_U\gets C_U\setminus N(w); L_V\gets C_V\setminus N(w)$\;
	\textbf{if} {$|L_U|<|L_V|$} \textbf{then} {$L_V\gets \emptyset$;}
	\textbf{else } {$L_U\gets \emptyset$;}
	
	lines~19-22 of Algorithm~\ref{alg:npivot}\;
	
	%move $w$ from $C_U$ (or $C_V$) into $P_U$ (or $P_V$)\;
		\npc($C_U, C_V, P_U, P_V, H_U, H_V$)
}
	
%	\Fn{\npc($C_U, C_V, P_U, P_V, H_U, H_V$)}{
%		lines~14-17 of Algorithm~\ref{alg:npivot}\;
%		
%		$w\gets \arg\min_{u\in C_U\cup C_V}{\left (\min(|C_U\setminus N(u)\setminus \{u\}|,|C_V\setminus N(u)\setminus \{u\}|)\right )}$\;
%	%	Let $w$ be the minimum non-neighbor node-pivot\;
%		$L_U\gets C_U\setminus N(w)\setminus \{u\}; L_V\gets C_V\setminus N(w)\setminus \{u\}$\;
%		\textbf{if} {$|L_U|<|L_V|$} \textbf{then} {$L_V\gets \emptyset$;}
%		\textbf{else } {$L_U\gets \emptyset$;}
%		
%		lines~19-23 of Algorithm~\ref{alg:npivot}\;
%		
%		%move $w$ from $C_U$ (or $C_V$) into $P_U$ (or $P_V$)\;
%	%	\npc($C_U, C_V, P_U, P_V, H_U, H_V$)
%	}
\end{algorithm}

\subsection{A Novel Minimum Non-neighbor Candidate Partition Strategy}\label{ssec:minimum}

%As a general framework, \npivot has no specific time complexity. We design a specific candidate partition strategy called minimum non-neighbor candidate partition. \npivot, when equips the minimum non-neighbor candidate partition, can get the best worst case time complexity in theorem.
As a general framework, \npivot does not inherently possess a specific time complexity. The time complexity depends on the specific design of the candidate partition strategy, i.e. how to choose $L_U$ and $L_V$ (line~18 of Algorithm~\ref{alg:npivot}). In this subsection, we propose the minimum non-neighbor candidate partition strategy, with which \npivot can achieve a lower worst-case time complexity than the previous SOTA works, as proven in Theorem~\ref{the:npc_complexity} and Theorem~\ref{the:all_complexity}.

We start with the definition of the minimum non-neighbor candidate partition.  The minimum non-neighbor candidate partition selects $L_U$ and $L_V$ based on the non-neighbors of the node with the fewest non-neighbors. The core idea behind the minimum non-neighbor candidate partition is to minimize the sizes of the sets $L_U$ and $L_V$ during the recursive calls in the \npc procedure. Each recursive node in the \npc search tree has $|L_U| + |L_V| + 1$ child nodes, so reducing the sizes of $L_U$ and $L_V$ leads to a smaller overall search tree and more efficient computation. 

\begin{definition}[Minimum non-neighbor candidate partition]\label{def:mncp}
	For the sets $C_U$ and $C_V$, the node with minimum non-neighbor is $$w\gets \arg\min_{u\in C_U\cup C_V}{\left (\min(|C_U\setminus N(u)|,|C_V\setminus N(u)|)\right )}.$$ 
	The partitioning is done as follows:
	\begin{equation}
		\begin{cases}
			L_U\gets C_U\setminus N(w)$, $L_V\gets \emptyset, \quad |C_U\setminus N(w)|\le |C_V\setminus N(w)| \\
			L_U\gets \emptyset, L_V\gets C_V\setminus N(w), \quad |C_U\setminus N(w)|>|C_V\setminus N(w)| \\
		\end{cases}
	\nonumber
	\end{equation}
\end{definition}

In Definition~\ref{def:mncp}, $w$ may be in $C_U$ or $C_V$. We mainly discuss the cases when \( w \in C_U \), and the following result also holds when \( w \in C_V \). When \( w \in C_U \), it is true that \( C_U \setminus N(w) = C_U \) because \( N(w) \) refers to the set of nodes on the opposite side of the bipartite graph connected to \( w \). If \( |C_U| = |C_U \setminus N(w)| \leq |C_V \setminus N(w)| \), we can conclude that \( L_U = C_U \), meaning that we can enumerate the nodes on the \( U \)-side directly. This condition suggests that when the size of \( C_U \) is small enough, it becomes efficient to enumerate the nodes in \( C_U \). On the other hand, if \( |C_U| = |C_U \setminus N(w)| \geq |C_V \setminus N(w)| \), the more efficient strategy is to enumerate the non-neighbors of \( w \) on the \( V \)-side, meaning \( L_V = C_V \setminus N(w) \). The partition optimizes the process by leveraging the smaller subset of nodes, ensuring that either side of the bipartite graph is efficiently handled depending on their relative sizes. Algorithm~\ref{alg:minnpivot} is the pseudo-code of the implementation with the minimum non-neighbor candidate partition strategy. The following example illustrates how our algorithm works.

\begin{figure*}[t]
	%	\vspace*{0.2cm}
	\begin{center}
		\includegraphics[width=0.8\linewidth]{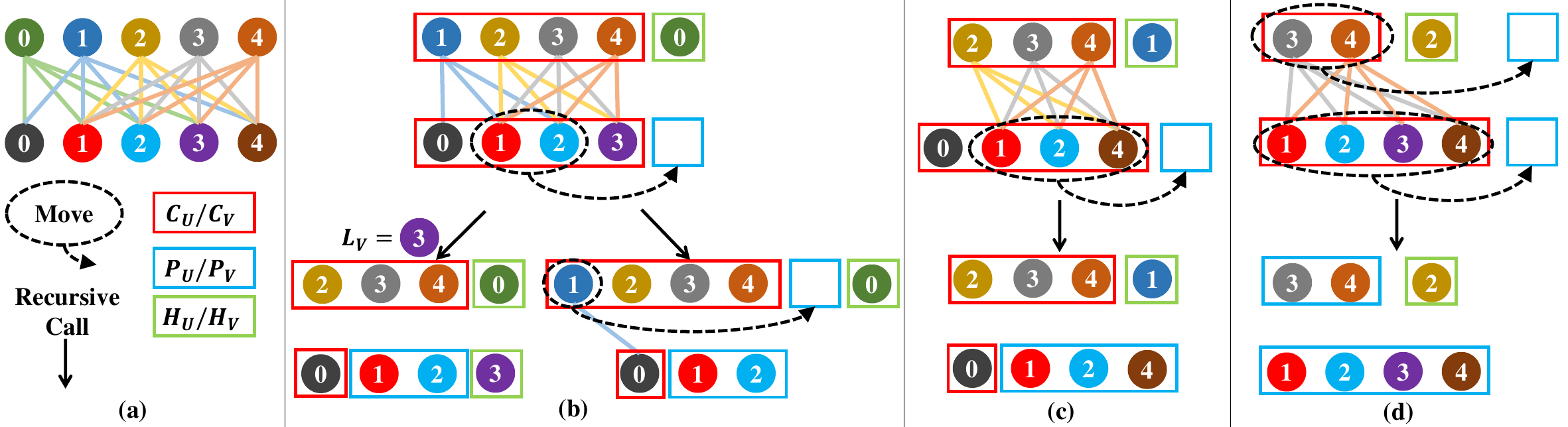}
	\end{center}
	%\vspace*{-0.2cm}
	\caption{A running example of our proposed \npivot }
	\label{fig:example}
	%\vspace*{-0.1cm}
\end{figure*}

\begin{example}\textit{
Figure~\ref{fig:example} illustrates the process of \ \npivot for counting $(3,3)$-biclique. The example graph and the associated icon description are presented in Figure~\ref{fig:example}(a). Figure~\ref{fig:example}(b), (c), and (d) are running processes of the \npc procedure, with the node-split strategy. For clarity, edges connected to the already selected nodes $P_U$, $P_V$, $H_U$, and $H_V$ (within the blue and green boxes) are omitted, as $(P_U \cup H_U, P_V \cup H_V)$ already forms a biclique, and $C_U$ and $C_V$ represent the sets of common neighbors of $P_U \cup H_U$ and $P_V \cup H_V$, respectively. The upper nodes are labeled from $u_0$ to $u_4$, and the lower nodes are labeled from $v_0$ to $v_4$. In Figure~\ref{fig:example}(b), \npivot counts the $(3,3)$-bicliques that contain $u_0$. So $u_0$ is in $H_U$. The candidate sets are now  $C_U=\{u_1,u_2,u_3,u_4\}$ and $C_V=\{v_0,v_1,v_2,v_3\}$. Since $v_1$ and $v_2$ connect to all nodes in $C_U$, they are moved into the node-pivots (represented by dotted cycles, moving from the red box to the blue box, as indicated in lines~16-17 of Algorithm~\ref{alg:npivot}). In the remaining candidate sets $C_U=\{u_1,u_2,u_3,u_4\}$ and $C_V=\{v_0,v_3\}$, $u_1$ has the minimum number of non-neighbors, and let $L_U=\emptyset, L_V=\{v_3\}$ (lines~3-5 of Algorithm~\ref{alg:minnpivot}). Put $v_3$ into $H_V$ and recursively call \npc to count the bicliques containing $v_3$ (line~23  of Algorithm~\ref{alg:npivot}). There is no edge between the candidate sets $C_U=\{u_2,u_3,u_4\}$ and $C_V=\{v_0\}$ (line~14 of Algorithm~\ref{alg:npivot}) and the \cnt procedure will add three $(3,3)$-bicliques. Then backtrack to the previous layer and recursively call the \npc procedure (line~23 of Algorithm~\ref{alg:npivot}). After moving $u_1$ into $P_U$, there is no edge between the candidate sets. In this case, there is no $L_U$ and $L_V$, and just recursively call the \npc procedure to count the result (line~23 and 14 of Algorithm~\ref{alg:npivot}). Figure~\ref{fig:example}(c) is running process of counting the $(3,3)$-bicliques that contain $u_1$, which has $3$ $(3,3)$-bicliques. Figure~\ref{fig:example}(d) is running process of counting the $(3,3)$-bicliques that contain $u_2$, which has $4$ $(3,3)$-bicliques.}
\end{example}

\stitle{Time Complexity of \npc and \npivot.} %Theorem~\ref{the:npc_complexity} and \ref{the:all_complexity} give the time complexity of the \npc procedure and \npivot framework with the minimum non-neighbor candidate partition.
The time complexity of the \npc procedure and the entire \npivot framework, when equipped with the minimum non-neighbor candidate partition strategy, is detailed in Theorem~\ref{the:npc_complexity} and Theorem~\ref{the:all_complexity}.

%\begin{theorem}
%	The minimum non-neighbor  node-pivot leads to the minimum value of $|L_U|+|L_V|+1$, i.e., the minimum number of recursive child nodes of a \npc search node.
%\end{theorem}

%\begin{theorem}
%	With the minimum non-neighbor node-pivot, the \npc procedure has time complexity of $O({n^2}3^{\frac{n}{3}})$ where $n=|C_U|+|C_V|$.
%\end{theorem}
%\begin{proof}
%\begin{equation}
%\begin{aligned}
%	T(n)& \le T(n-k-1)+\sum_{i=1}^{k}{T(n-k-i)}
%	\\ & \le (k+1)T(n-k-1)
%\end{aligned}
%\end{equation}
%\end{proof}
\begin{theorem}\label{the:npc_complexity}
	 With the minimum non-neighbor candidate partition strategy, the \npc procedure has time complexity of $O(2^{\frac{n'}{2}})$ where $n'=|C_U|+|C_V|$.
\end{theorem}
\begin{proof}
	 The time complexity of  finding the node-pivot in each recursive search vertex of \npc is $O({n'}^2)$, which is equal to compute the degree of each node. For simplicity, we use $n$ instead of $n'$ in the following context.
	
	Let $T(n)$ be the time complexity of \npc with input $C_U, C_V$. Let $k$ be the number of the minimum non-neighbors and $c_3 $ be constant. We can derive that 
\begin{equation}\label{equ:complexityStart}
		T(n)\le\sum_{i=1}^{k}T(n-k-i) + T(n-k-1) + c_3n^2,
	\end{equation}
	where $\sum_{i=1}^{k}T(n-k-i)$ is the time complexity of lines~19-22 of Algorithm~\ref{alg:npivot}, $T(n-k-1)$ is the time complexity of line~23 of Algorithm~\ref{alg:npivot} and $c_3n^2$ is the time complexity of finding the node-pivots. 
Define constant numbers $c_1,c_2,c_3$ that satisfy $c_1\ge 4.5(c_2+c_3)$. We can prove that
\begin{equation}\label{equ:res}
		T(n)\le c_1 2^{n/2}-c_2 n^2.
	\end{equation}
	
	When $n=1$, we have $T(1)\le c_1\sqrt{2}-c_2$ from inequality~(\ref{equ:res}) and $T(1)\le  c_3$ from inequality~(\ref{equ:complexityStart}). inequality~(\ref{equ:res}) holds because $c_1\ge 4.5(c_2+c_3) > \frac{c_2+c_3}{\sqrt{2}}$. 	
	
	Let $N$ be a positive integer. We assume that inequality~(\ref{equ:res}) holds for all integers $n\in [1, N-1]$, and prove that it also holds for $n=N$.
	
	\stitle{Case $k=2$.}  inequality~(\ref{equ:complexityStart}) becomes $T(n)\le 2T(n-3)+T(n-4) + c_3n^2$. According to the assumption, we have  
	\begin{equation}
		\begin{aligned}
			T(n) &\le  2T(n-3)+T(n-4)+ c_3n^2
			\\ & \le  2c_12^{\frac{n-3}{2}} -c_2(n-3)^2 +c_12^{\frac{n-4}{2}} -c_2(n-4)^2+ c_3n^2
			\\ & \le  (2^{-\frac{1}{2}}+2^{-2})c_1 2^{\frac{n}{2}}  + c_3n^2
		%	\\ & = (2^{-\frac{1}{2}}+2^{-2})c_1 2^{\frac{n}{2}} +(c_3-2c_2)n^2+14c_2n-27c_2
			\\ & \le c_1 2^{\frac{n}{2}} - c_2 n^2.
		\end{aligned}
	\end{equation}
The last inequality equals to prove
\begin{equation}
	\frac{n^2}{2^{n/2}}\le \frac{(1-2^{-\frac{1}{2}}-2^{-2})c_1}{c_2+c_3}\le \frac{c_1}{c_2+c_3}
\end{equation}
holds for all $n$. Since $\frac{n^2}{2^{n/2}} \le 4.5$ ($\frac{n^2}{2^{n/2}} =4.5$ when $n=6$) and $\frac{c_1}{c_2+c_3}\ge 4.5$, the mathematical induction is proved.
	
	\stitle{Case $k\neq 2$.} A scaled inequality~(\ref{equ:complexityStart}) is
	\begin{equation}\label{equ:complexityTwo}
		T(n)\le (k+1)T(n-(k+1)) + c_3n^2, 
	\end{equation}
	from which we have
	\begin{equation}
		\begin{aligned}
		T(n) &\le  (k+1) c_1 2^{(n-(k+1))/2} - c_2(k+1)(n-(k+1))^2 + c_3n^2
			\\ & \le  (k+1)2^{-\frac{k+1}{2}} \cdot c_12^{n/2} +  c_3n^2
			\\ & \le c_1 2^{\frac{n}{2}} - c_2 n^2.
		\end{aligned}
	\end{equation}

	The last inequality comes from (1) $(k+1)2^{-\frac{k+1}{2}}$ is always no larger than $1$; (2) the following inequality is always true
	\begin{equation}
		\frac{n^2}{2^{n/2}}\le \frac{(1-(k+1)2^{-\frac{k+1}{2}})c_1}{c_2+c_3} \le \frac{c_1}{c_2+c_3}
	\end{equation}
Since $\frac{n^2}{2^{n/2}} \le 4.5$ ($\frac{n^2}{2^{n/2}} =4.5$ when $n=6$) and $\frac{c_1}{c_2+c_3}\ge 4.5$, the mathematical induction is proved.

As a result, the time complexity of \npc  is $O(2^{n'/2})$.
\end{proof}

\begin{theorem}\label{the:all_complexity}
	With the minimum non-neighbor candidate partition strategy, the time complexity of  Algorithm~\ref{alg:npivot} is $O(|E|2^{n_{max}/2})$, where $n_{max} = \max_{u\in U}{|N(u, V)|} + \max_{v\in V}{|N(v, U)|}$.
\end{theorem}
\begin{proof}
	The number of calling \npc is bounded by $|E|$, and the size of $|C_U|+|C_V|$ is bounded by $n_{max}$. Thus, according to Theorem~\ref{the:npc_complexity}, the time complexity of Algorithm~\ref{alg:npivot} is $O(|E|2^{n_{max}/2})$.
\end{proof}

%\begin{table}[h!]
%	%	\scriptsize
%	\centering
%	\caption{Time complexity  ($2^{\frac{1}{2}}\approx1.414, 3^{\frac{1}{3}}\approx 1.442$)}
%	%	\vspace*{-0.3cm}
%	% \centering
%	\begin{tabular}{c|c|c}
%		\toprule
%		\npivot & \edgepivot & \bc \\
%		%		\cmidrule{1-3}
%		\midrule
%		$O(|E|2^{n_{max}/2})$ & $O(|E|3^{|E_{max}|/3})$  & $O(\alpha(H)^{p-2} |E'| d_{max})$ \\
%		\bottomrule
%	\end{tabular}
%	%\vspace*{-0.3cm}
%	\label{tab:complexity}
%\end{table}

\begin{table}[t]
	%	\scriptsize
	\centering
	\caption{Time complexity comparison  ($2^{\frac{1}{2}}\approx1.414, 3^{\frac{1}{3}}\approx 1.442$)}
	%	\vspace*{-0.3cm}
	% \centering
	\begin{tabular}{c|c|c}
		\toprule
		\npivot (Our) & \edgepivot \cite{edgepivot}& \bc \cite{BClistYangPZ21} \\
		%		\cmidrule{1-3}
		\midrule
		$O(|E|2^{n_{max}/2})$ & $O(|E|3^{|E_{max}|/3})$  & $O(\alpha(H)^{p-2} |E'| d_{max}+\triangle)$ \\
		\bottomrule
	\end{tabular}
	%\vspace*{-0.3cm}
	\label{tab:complexity}
\end{table}

\stitle{The Superiority of \npivot on the Time Complexity.} Table~\ref{tab:complexity} summarizes the time complexity of the proposed \npivot and SOTA methods. The time complexity of \edgepivot is $O(|E|3^{|E_{max}|/3})$, where $|E_{max}|$ is bounded by the maximum number of edge among all candidate sets of edge-split strategy.   Since $2^{\frac{1}{2}}\approx1.414, 3^{\frac{1}{3}}\approx 1.442$ and $|E_{max}|> n_{max}$, our \npivot has a lower worst-case time complexity. The time complexity of \bc is \(O(\alpha(H)^{p-2} |E'| d_{\text{max}}+\triangle)\), as outlined in Section~\ref{ssec:bc}. Here, \(\alpha(H)\) represents the  arboricity in the 2-hop graph \(H\), and in the worst case, \(\alpha(H)\) is on the order of \(n_{\text{max}}\) \cite{BClistYangPZ21, arboricity, Parboricity}.  $|E'|$ is in the order of $d_{max}|E|$. $\triangle$ is the result size, which can be quite large for large real-world networks. Therefore, \npivot also outperforms \bc.

\subsection{The Proposed Cost Estimator}\label{ssec:cost_estimator}

\begin{algorithm}[t]
	\caption{Cost estimator (line~4 of Algorithm~\ref{alg:npivot})}
	\label{alg:cost_estimator}
	\small
	%\scriptsize
	\KwIn{The graph $G(U, V, E)$,  a node $u$, two integers $x, y$}
	\KwOut{True if edge-split, False if node-split}
	\SetKwProg{Fn}{Procedure}{}{}
	\Fn{$\kw{costEs}(l,r,e)$}{
		\textbf{if} {$l<x$ or $r<y$} \textbf{then} {\Return{$0$}\;} 
		\Return{$((\frac{e}{\min(l,r)})^{min(l,r)}, 2^{\frac{\min(l,r)}{2}})$\;}
	}
	Denote $R$ by the rank of nodes\;
	$cost_{node}\gets 0; cost_{edge} \gets 0; l\gets 0; r\gets 0; e \gets 0$\;
	$cnt_w\gets 0, \forall w\in \cup_{v\in N(u)}{N(v)}$\;
	%	$l\gets 0$\;
	\For{$v\in N(u)$}{
		\For{$w\in N(v)$ s.t. $R(w)>R(u)$}{
			$cnt_w\gets cnt_w+1$\;
			%\If{$cnt_w=y$}{$l\gets l+1$;}
			\textbf{if} {$cnt_w=y$} \textbf{then} {$l\gets l+1$\;}
		}
	}
	
	%	$ r\gets 0$\;
	Label  $N(u)$ by $\{v_1,v_1,v_2,...v_{d(u)}\}$ with $R(v_i)<R(v_{i+1})$\;
	$ss\gets \{0,0,...0\}$; \textit{/*suffix sum vector with length ${d(u)}$ */}\\
	\For{$i=1$ to ${d(u)}$} {
		$l'_w\gets 0$\;
		\For{$w\in N(v_i)$ s.t. $R(w)>R(u)$}{
			%\If{$cnt_w\ge y$}{$l'_v=l'_v+1$;}
			\textbf{if} {$cnt_w\ge y$} \textbf{then} {$\{l'_w=l'_w+1; e\gets e + 1\}$\;}
		}
		\textbf{if} {$l'_w\ge x-1$} \textbf{then} {$\{r\gets r+1$; $ss_i\gets 1\}$\;}
	}

	%\For{$i=n$ to $2$}{$ss_{i-1}\gets ss_{i-1}+ss_{i}$;}
	\textbf{for} {$i={d(u)}$ to $2$} \textbf{do} {$ss_{i-1}\gets ss_{i-1}+ss_{i}$\;}
	
	\For{$i=1$ to ${d(u)}$}{
		%\For{$v\in N(u)$}{
			$l'\gets 0$; $r'\gets ss_i$; $e'\gets 0$\;
			\For{$w\in N(v_i)$ s.t. $R(w)>R(u)$} {
				%\If{$cnt_w\ge y$}{$l'_v=l'_v+1$;}
				\textbf{if} {$cnt_w\ge y$} \textbf{then} {$\{l'=l'+1;  e'\gets e' + cnt_w\}$\;}
				$cnt_w\gets cnt_w-1$\;
			}
			%		\textbf{if} {$l'_w\ge x-1$} \textbf{then} {$ cost_{edge}\gets  cost_{edge} + 2^{(l'_w+ss_i)/2}$\;}
			\textbf{if} {$l'\ge x-1$} \textbf{then} {$ cost_{edge}\gets  cost_{edge} + \kw{costEs}(l', r', e')$\;}
		}
		
		$cost_{node}\gets \kw{costEs}(l,r,e)$\;

		\textbf{if} {$cost_{node}< cost_{edge}$} \textbf{then} {\Return False;} \textbf{else} {\Return True\;}
		%	}
\end{algorithm}

%\stitle{Intuitive difference between the two graph-split strategies.} \bc use the node-split strategy, and \edgepivot use the edge-split strategy (Section~\ref{sec:existingworks}). Intuitively, node-split is more efficient than edge-split on dense graphs, and edge-split is more efficient than node-split on sparse graphs. On dense graphs, the subgraphs of the edge-split may be near the whole graph. On the sparse graphs, the subgraphs of edge-split is smaller than the whole graph. 

%\stitle{Intuitive Difference Between the Two Graph-Split Strategies.} 
The \bc algorithm employs a node-split strategy, while \edgepivot uses an edge-split strategy (Section~\ref{sec:existingworks}). Intuitively, the node-split strategy tends to be more efficient for dense graphs with more connections. In such graphs, the subgraphs generated from edge-splitting can be nearly as large as the entire graph, resulting in minimal reduction in problem size and inefficiency. Conversely, in sparse graphs, the edge-split strategy becomes more efficient. In this case, the subgraphs created by splitting edges are much smaller, allowing \edgepivot to perform faster due to the reduced search space. Thus, node-split excels in handling dense structures, whereas edge-split is more suited for sparse graph scenarios.

%In this section, we provide a cost estimator to choose the graphs split strategy (line~4  of Algorithm~\ref{alg:npivot}).  The cost estimator is responsible for judging whether node-split (lines~10-11 of Algorithm~\ref{alg:npivot}) or edge-split (lines~5-8 of Algorithm~\ref{alg:npivot}) is more efficient. 

%In this subsection, we present a cost estimator that helps determine whether to use the node-split (lines~10-11) or edge-split (lines~5-8) strategy in line~4 of Algorithm~\ref{alg:npivot}. 

To leverage the advantages of both the node-split and edge-split strategies, we propose a novel cost estimator that helps determine when to use the node-split (lines 10-11) or edge-split (lines 5-8) strategy in line~4 of Algorithm~\ref{alg:npivot}. The goal of this estimator is to predict which graph-split strategy will be more efficient. Specifically, let the variables $l$, $r$, and $e$ be the sizes of the candidate sets and the number of edges as follows:
\( l = |C_U| \) (the size of the candidate set $C_U$), \( r = |C_V| \) (the size of the candidate set $C_V$), \( e = |(C_U \times C_V) \cap E| \) (the number of edges between $C_U$ and $C_V$). Theorem~\ref{the:npc_complexity} initially suggests using \( 2^{\frac{l+r}{2}} \) as a measure of the running cost for the recursive process. However, this measure has two main drawbacks:
\textbf{1. Imbalance Between $l$ and $r$}: When the sizes of $l$ and $r$ differ significantly, the smaller value dominates the computation cost, rather than their sum. To account for this, we replace \( 2^{\frac{l+r}{2}} \) with \( 2^{\frac{\min(l,r)}{2}} \), focusing on the smaller of the two sets.
\textbf{2. Edge Density Consideration}: The initial measure only accounts for the number of nodes and ignores the number of edges, which can significantly impact the algorithm's performance. When the number of edges $e$ is small, the algorithm can still run efficiently even if the combined size of $l$ and $r$ is large.

To address these limitations, we propose a new metric that incorporates both the number of nodes and edges. This new metric is based on the observation that the complexity of the algorithm increases as we process more edges between $C_U$ and $C_V$. The refined metric is defined as follows:
\begin{equation}  
\left( \frac{e}{\min(l,r)} \right)^{\min(l,r)}.
\end{equation}

This term reflects the idea that if a set contains \( a \) items, the number of subsets of size \( b \) is proportional to \( {a \choose b} \), which is bounded by \( a^b \). In this case, \( a = \frac{e}{\min(l,r)} \) and \( b = \min(l,r) \).
Here, \( \frac{e}{\min(l,r)} \) represents the average number of neighbors for a node in the smaller of the two sets, while \( \left( \frac{e}{\min(l,r)} \right)^{\min(l,r)} \) estimates the complexity of processing all possible subsets of neighbors of size $\min(l,r)$. The final running cost estimation of the \npc procedure combines both node and edge considerations:
\begin{equation}
\min \left( \left( \frac{e}{\min(l, r)} \right)^{\min(l,r)}, 2^{\frac{\min(l, r)}{2}} \right)
\end{equation}
This metric balances the trade-offs between node size and edge density, effectively estimating the computational cost. As demonstrated in our experiments, this cost estimation provides reliable guidance for choosing the most efficient graph-split strategy.

%Let $l,r,e$ represents $|C_U|,|C_V|,|(C_U\times C_V) \cap E|$ respectively. Theorem~\ref{the:npc_complexity} suggests to let $2^{\frac{l+r}{2}}$ be a measure of running cost. This measure has two drawbacks. The first drawback is that when $l$ and $r$ differs a lot, $min(l,r)$ dominates the cost instead of $l+r$. Thus, we use $2^{\frac{\min(l,r)}{2}}$ instead of  $2^{\frac{l+r}{2}}$ as a measure. Second, this measure only takes into account the number of nodes. In practice, the number of edges also affects the performance. When the number of edges is small, the algorithm can run fast even if $l+r$ is large. Therefore, we also consider a new metric $(\frac{e}{\min(l, r)})^{min(l,r)}$. Given a set of $a$ items, the number of size $b$ subsets is in the order of ${a \choose b}<a^b$. In this metric, $a=\frac{e}{\min(l, r)}$ and $b=min(l,r)$ . $\frac{e}{\min(l, r)}$ represents the average number of neighbors and $(\frac{e}{\min(l, r)})^{min(l,r)}$ represents the order of the size $min(l,r)$ subsets of the neighbors. At last, the running cost estimation of the \npc procedure is $$\min((\frac{e}{\min(l, r)})^{\min(l,r)}, 2^{\frac{\min(l, r)}{2}}).$$ As shown in the experiments, the cost estimation performs well.

The cost estimator computes two values $cost_{node}$ and $cost_{edge}$, which represent the computation cost of the two strategies respectively. For $cost_{node}$, we compute the value of $l,r,e$ and let $cost_{node} = \min((\frac{e}{\min(l, r)})^{\min(l,r)}, 2^{\frac{\min(l, r)}{2}})$. The computation of $cost_{edge}$ is more complex. In Algorithm~\ref{alg:npivot}, the edge-split has multiple calls of \npc (line~8). Hence, we compute cost estimation for each edge and sum them together as the final value of $cost_{edge}$.

Algorithm~\ref{alg:cost_estimator} outlines the pseudo-code of the cost estimator. Algorithm~\ref{alg:cost_estimator} inputs a node $u$ and two integers $x, y$, and returns whether node-split or edge-split is more efficient to count $(p,q)$-biclique with $p\ge x, q\ge y$. The \kw{costEs} procedure (lines~1-3 of Algorithm~\ref{alg:cost_estimator}) returns the cost estimation of a \npc procedure with input $l=|C_U|, r=|C_V|, e=|(C_U\times C_V)\cap E|$.  For approximating $cost_{node}$, Algorithm~\ref{alg:cost_estimator} maintains $l,r,e$ (line~5).  $l$ is the number of nodes in $U$ that have larger rank than $u$ and has more than $y$ neighbors in $N(u)$ (lines~7-10). $r$ is the number of nodes in $N(u)$ that have more than $x-1$ neighbors in the previous $l$ nodes (lines~13-17). $e$ is the number of edges between the $l$ and $r$ nodes (lines~13-16). At last, $cost_{node}=\min((\frac{e}{\min(l, r)})^{\min(l,r)}, 2^{\frac{l+r}{2}})$ (line~25). For approximating $cost_{edge}$, Algorithm~\ref{alg:cost_estimator} approximates $l',r',e'$ for each edge (line~19-24). Consider an edge $(u, v_i)$.  $l'$ is the number of nodes in $U$ that has larger rank than $u$ and has more than $y$ neighbors in $S_{u,v_i}=\{v|v\in N(u), R(v)>R(v_i)\}$ (lines~19-22).  $r'$ is the number of nodes in $S_{u,v_i}$ that has more than $x-1$ neighbors (line~, 17, 18, 20). Label the nodes in $N(u)$ by $\{v_1,v_1,v_2,...v_{d(u)}\}$ (line~11). Instead of computing $r'$ for each edge,  Algorithm~\ref{alg:cost_estimator} records a suffix summary vector $ss$ where $ss_i$ is the number of nodes in $\{v|v\in N(u), R(v)>R(v_i)\}$ that has more than $x-1$ neighbors (lines~17-18). Thus, we can use $ss_i$ to approximate $r'$ directly (line~20). Due to the edges before $v_i$ being removed already (line~23), $cnt_w$ is the number of neighbors of $w$  with rank larger than $v_i$. Thus, sum all the $sum_w$ together is the number of edges $e'$ (line~22). At last, $cost_{edge}=\min((\frac{e'}{\min(l', r')})^{\min(l',r')}, 2^{\frac{l'+r'}{2}})$ (line~24).

\begin{table*}[t!]
	\small
	%	\scriptsize
	\centering
	\caption{Datasets. $\bm{\overline{d}_U}$ and $\bm{\overline{d}_V}$ are  the average degrees for $U$ and $V$, respectively.}
	%	\vspace*{-0.3cm}
	\begin{tabular}{c|c|c|c|c|c|c|c|c|c}
		\toprule
		\textbf{Datasets} & \textbf{Abbr} & $\bm{|U|}$ & $\bm{U}$ \textbf{Type} & $\bm{|V|}$ & $\bm{V}$ \textbf{Type} & $\bm{|E|}$ & $\bm{E}$ \textbf{Type} & $\bm{\overline{d}_U}$ & $\bm{\overline{d}_V}$\\
		\midrule
		
		\textbf{Youtube}  & You
		& 94,238  & User  & 30,087  & Group
		& 293,360  & Membership
		& 3.11   & 9.75  \\
		\textbf{DailyKos}  & Kos
		& 3,430  & Document  & 6,906  & Word
		& 353,160  & Occurrence
		& 102.96   & 51.14  \\
		\textbf{Bookcrossing}  & Book
		& 77,802  & User  & 185,955  & Book
		& 433,652  & Rating
		& 5.57   & 2.33  \\
		\textbf{Github}  & Git
		& 56,519  & User  & 120,867  & Project
		& 440,237  & Membership
		& 7.79   & 3.64  \\
		\textbf{Citeseer}  & Cite
		& 105,353  & Author  & 181,395  & Publication
		& 512,267  & Authorship
		& 4.86   & 2.82  \\
		\textbf{Stackoverflow}  & Stac
		& 545,195  & User  & 96,678  & Post
		& 1,301,942  & Favorite
		& 2.39   & 13.47  \\
		\textbf{Twitter}  & Twit
		& 175,214  & User  & 530,418  & Hashtag
		& 1,890,661  & Usage
		& 10.79   & 3.56  \\
		\textbf{IMDB}  & IMDB
		& 685,568  & Person  & 186,414  & Work
		& 2,715,604  & Association
		& 3.96   & 14.57  \\
		\textbf{DiscogsGenre}  & Genre
		& 1,754,823  & Artist  & 15  & Genre
		& 3,142,059  & Feature
		& 1.79   & 209470.60  \\
		\textbf{Actor2}  & Act
		& 303,617  & Movie  & 896,302  & Actor
		& 3,782,463  & Appearance
		& 12.46   & 4.22  \\
		\textbf{Amazon}  & Ama
		& 2,146,057  & User  & 1,230,915  & Item
		& 5,743,258  & Rating
		& 2.68   & 4.67  \\
		\textbf{DBLP}  & DBLP
		& 1,953,085  & Author  & 5,624,219  & Publication
		& 12,282,059  & Authorship 
		& 6.29   & 2.18  \\
		\bottomrule
	\end{tabular}
	%	\vspace*{-0.4cm}
	\label{tab:datasets}
\end{table*}

\begin{theorem}\label{the:buildindextime}
Given a bipartite graph $G(U,V,E)$, in Algorithm~\ref{alg:npivot} (line~4), the total time complexity of Algorithm~\ref{alg:cost_estimator} is $O(\alpha(G)|E|)$, where $\alpha(G)$ is the arboricity of $G$.
\end{theorem}
\begin{proof}
	For each node $u$ (i.e., line~3 of Algorithm~\ref{alg:npivot}), lines~7, 13 and 19 of  Algorithm~\ref{alg:cost_estimator} scan the neighbors of $u$, which has complexity of $O(|E|)$. Each line of lines~8, 15, and 21 of  Algorithm~\ref{alg:cost_estimator} scan the neighbors of a neighbor of $u$, which has a total time complexity of $O( \sum_{(u,v)\in E}d(v))$. According to the definition of $\alpha(G)$  \cite{CN}, $E$ can be partitioned into $\alpha(G)$ edge-disjoint spanning forests, and let these trees be $F_i, 1\le i\le \alpha(G)$, where $|F_i|\le |V|$. We have $\sum_{(u,v)\in E}d(v) = \sum_{i}^{\alpha(G)}\sum_{(u,v)\in F_i}{d(v)}\le \sum_{i}^{\alpha(G)}\sum_{v\in V}{d(v)}=2\alpha(G)|E|$. Thus, the complexity of  Algorithm~\ref{alg:cost_estimator} is $O(\alpha(G)|E|)$.
\end{proof}

%Theorem~\ref{the:buildindextime} shows that cost estimator has poly-nominal time complexity, which is very small compared to the counting problem.  Experiments show that with the cost estimator, our algorithm is more efficient than  using only one fixed graph-split strategy.

Theorem~\ref{the:buildindextime} demonstrates that the cost estimator has a polynomial time complexity, which is significantly smaller compared to the complexity of the biclique counting algorithm. By efficiently estimating the cost of various candidate sets, the cost estimator allows the algorithm to choose the most optimal strategy dynamically. Our empirical results  confirm that the proposed algorithm, when equipped with the cost estimator, outperforms approaches relying on a single fixed graph-split strategy (Section \ref{sec:exp}).

\stitle{Cost Estimator based Index.} %When counting $(p,q)$-biclique, it is clear that the parameters $x$ and $y$ should be no larger than $p$ and $q$ respectively. In practice, we build an index for the integer $x$ and $y$. The index is a vector with length $|U|$ that save the results returned by Algorithm~\ref{alg:cost_estimator} for all nodes in $U$. With the help of the index, when $p\ge x, q\ge y$, Algorithm~\ref{alg:npivot} costs only linear time to judge whether node-split or edge-split is more efficient. 
When counting $(p,q)$-bicliques, the parameters $x$ and $y$ should not exceed $p$ and $q$. To optimize this process, we can build an index for $x$ and $y$. Specifically, the index is a vector of length $|U|$, storing the results of Algorithm~\ref{alg:cost_estimator} for each node in $U$. Equipped with the pre-computed index, when $p \geq x$ and $q \geq y$, Algorithm~\ref{alg:npivot} can determine whether a node-split or edge-split is more efficient in constant time complexity for each node. The time complexity of building the index is Theorem~\ref{the:buildindextime}.

\section{ Further  Optimizations}\label{sec:opt}
In this section, we devise several optimization tricks to further improve the efficiency of our proposed algorithm.

\stitle{\((\alpha, \beta)\)-core Based Graph Reduction \cite{BClistYangPZ21}.} For the bipartite graph $G(U,V,E)$ and any \((p,q)\)-biclique $(X,Y)$, every node \(u\in X\) must have at least \(q\) neighbors in \(V\), and every node \(v \in Y\) must have at least \(p\) neighbors in \(U\). This allows us to simplify the graph by reducing it to its \((p,q)\)-core, which retains only the nodes and edges that are essential for potential \((p,q)\)-bicliques \cite{BClistYangPZ21}. The most efficient algorithm for computing the \((p,q)\)-core of a bipartite graph \(G(U,V,E)\) operates in linear time with a complexity of \(O(|E|)\) \cite{alphabetacoreCIKM}. 

\begin{definition}[$(p,q)$-core \cite{alphabetacoreCIKM,alphabetacoreWWW}]\label{def:pqcore}
	For a bipartite graph \(G(U,V,E)\) and two integers \(p\) and \(q\), the \((p,q)\)-core of \(G\) is a maximal subgraph \(G_c(U_c,V_c,E_c)\) where every node in \(U_c\) has at least \(q\) neighbors in \(V_c\), and every node in \(V_c\) has at least \(p\) neighbors in \(U_c\).
\end{definition}

%\stitle{core value-based graph ranking.} The core ordering \cite{corePeel} is a widely adopted graph ordering strategy for graph mining problems\cite{PIVOTER, BClistYangPZ21, kclistplusplus, EppsteinLS13}. The core value of a node $u$ is the maximum degree of $u$ among the subgraphs that $u$ has the smallest degree in this subgraph \cite{subgraphcountingbook}. We utilize the core value-based ranking, where nodes are ranked by their core values.  The larger rank nodes have larger core values. This ranking makes sure only the larger core value nodes are in the candidate sets (lines~6, 7, 10 of Algorithm~\ref{alg:npivot}).

\stitle{Core Value-Based Node Ranking.} The core ordering method \cite{corePeel} is a commonly used strategy in graph mining tasks to enhance the algorithm efficiency \cite{PIVOTER, BClistYangPZ21, kclistplusplus, EppsteinLS13}. The core value of a node \(u\) is defined as the highest degree that \(u\) holds in a subgraph where \(u\) has the smallest degree within that subgraph \cite{subgraphcountingbook}. In this optimization, nodes are ranked according to their core values, where higher-ranked nodes correspond to those with larger core values. For any node, the number of the nodes with larger core values is bounded by the \textit{degeneracy} of the graph \cite{subgraphcountingbook}, which is often a small number for real-world networks \cite{ronghuaIO}. This ranking ensures that only nodes with larger core values are included in the candidate sets during key steps of the \npc procedure (lines 6, 7, 10 of Algorithm~\ref{alg:npivot}), optimizing the search space and improving computational performance. %\textcolor{blue}{. degeneracy is bound and thus the size of DAG is bound} %建议多说点

\stitle{Early Termination.} The \npc procedure can be terminated early under the following conditions: 

(1) Lower bounds on biclique size: If \(|H_U| > p\) or \(|H_V| > q\), the biclique is guaranteed to exceed the desired \((p, q)\)-biclique size, as the nodes in \(H_U\) and \(H_V\) must be part of the solution (Theorem~\ref{the:combination}).

(2) Upper bounds on biclique size: If \(|C_U \cup P_U \cup H_U| < p\) or \(|C_V \cup P_V \cup H_V| < q\), the search can be terminated since these sets represent the upper bounds on the biclique size, making it impossible to reach a valid \((p, q)\)-biclique.

(3) Single node in \(C_U\): When \(|C_U| = 1\), meaning that \(C_U\) contains only one node \(u\), instead of branching into further recursive calls, the biclique count can be computed directly as \({p_0 \choose p - h_0}{p_1 + c_1 \choose q - h_1} + {p_0 \choose p - h_0 - 1}{p_1 + n_u \choose q - h_1}\), where \(n_u = |N(u) \cap C_V|\). The first term represents the bicliques that do not include \(u\), while the second term counts bicliques containing \(u\).

(4) Single node in \(C_V\): Similarly, when \(|C_V| = 1\) (i.e., \(C_V\) contains only one node \(v\)), the biclique count is \({p_0 + n_v \choose p - h_0}{p_1 \choose q - h_1 - 1} + {p_0 + c_0 \choose p - h_0}{p_1 \choose q - h_1}\). The first term corresponds to bicliques that include \(v\), and the second term counts bicliques that do not include \(v\).

\stitle{Efficient Maintenance of the Non-neighbor Count.} During the \npc procedure in Algorithm~\ref{alg:npivot}, instead of recalculating the non-neighbor count from scratch at each recursive step, the algorithm reuses the non-neighbor counts from the parent search vertex. Specifically, for a node \(w \in C_U\), let \(C'_U = C_U \setminus \{w\}\). The non-neighbor set of any neighbor \(u \in N(w, C_V)\) remains unchanged, i.e., \(C_U \setminus N(u) = C'_U \setminus N(u)\). For non-neighbors \(u \in C_V \setminus N(w)\), the non-neighbor set decreases by exactly one, i.e., \(C_U \setminus N(u) \setminus \{w\} = C'_U \setminus N(u)\). Leveraging this observation, the optimization effectively manages and updates non-neighbor counts without redundant recalculations.

\section{Experiments}\label{sec:exp}

We conduct extensive experiments to answer the following Research Questions. \textbf{RQ1:} How fast our algorithm runs? \textbf{RQ2:} What is the effectiveness of the proposed cost estimator? \textbf{RQ3:} What is the effect of the graph density, $p$  and $q$ on the efficiency of our \npivot?   \textbf{RQ4:} How fast the running time of local counting and range counting? 

\subsection{Experimental Setup}

We evaluate all algorithms on a server with an AMD 3990X CPU and 256GB memory running Linux CentOS 7 operating system. We run algorithms against a single core. We terminate an algorithm if the running time is more than $10^6$ mile-seconds (ms).

\stitle{Datasets.} We evaluate our algorithms on 12 real-world bipartite networks drawn from various domains, covering a wide range of applications and structural characteristics. These datasets, listed in Table~\ref{tab:datasets}, are downloaded from the KONECT project (\url{http://konect.cc/networks/}), which compiles diverse network datasets for research purposes. These datasets vary widely in terms of size and density, with node counts ranging from thousands to millions, and edge counts reflecting the number of connections between the two sets of entities. The average degrees for both sets of nodes, denoted as $\overline{d}_U$ and $\overline{d}_V$, provide insight into the density of graph, highlighting the diversity in structure across different domains. These characteristics make the datasets an excellent test bed for evaluating the performance of biclique counting algorithms \cite{edgepivot,BClistYangPZ21}.

%All algorithms input the bipartite graph $G$, the biclique size $p$ and $q$ as the parameters.

\stitle{Algorithms.} We compare \bc \cite{BClistYangPZ21} and \edgepivot \cite{edgepivot} with our \npivot.  Both \edgepivot and \npivot apply to the range counting problem. \npivot can also locally count  the $(p,q)$-biclique for each node. All the algorithms are implemented in C++. The code of \bc and \edgepivot are from the open-source repository \cite{BClistYangPZ21, edgepivot}.  In default, \npivot uses the cost estimator index with $x=y=\min(p,q)$. The index building time is not included in our \npivot because the computation time of the cost estimator index is negligible (compared to the biclique counting time) as shown in Exp-5, and the index creation can be treated as a preprocessing step.   To be more reliable,
we  report the
average running time and quality of 10 runs per algorithm.

\begin{figure*}[t!]
	%	\vspace*{0.2cm}
		\begin{center}
\subfigure[You]{\label{sfig:time_you}\includegraphics[width=0.3\linewidth]{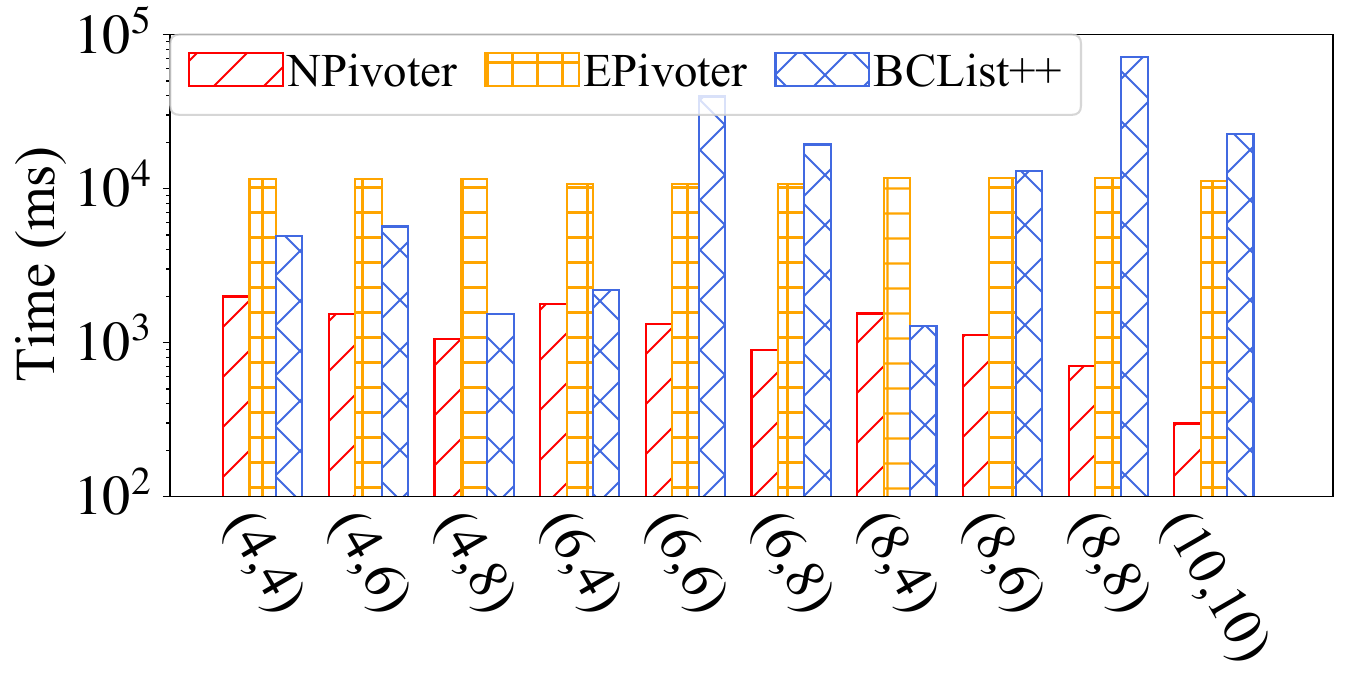}}
\subfigure[Kos]{\label{sfig:time_kos}\includegraphics[width=0.3\linewidth]{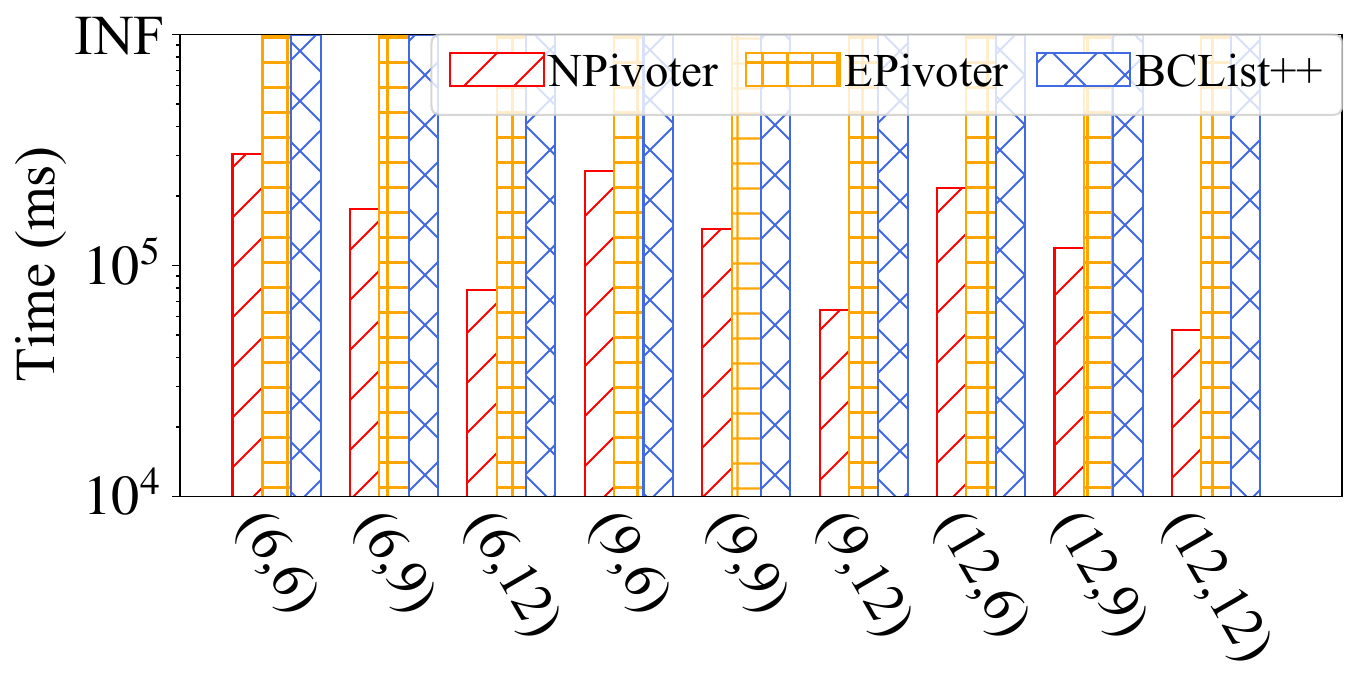}}
\subfigure[Book]{\label{sfig:time_book}\includegraphics[width=0.3\linewidth]{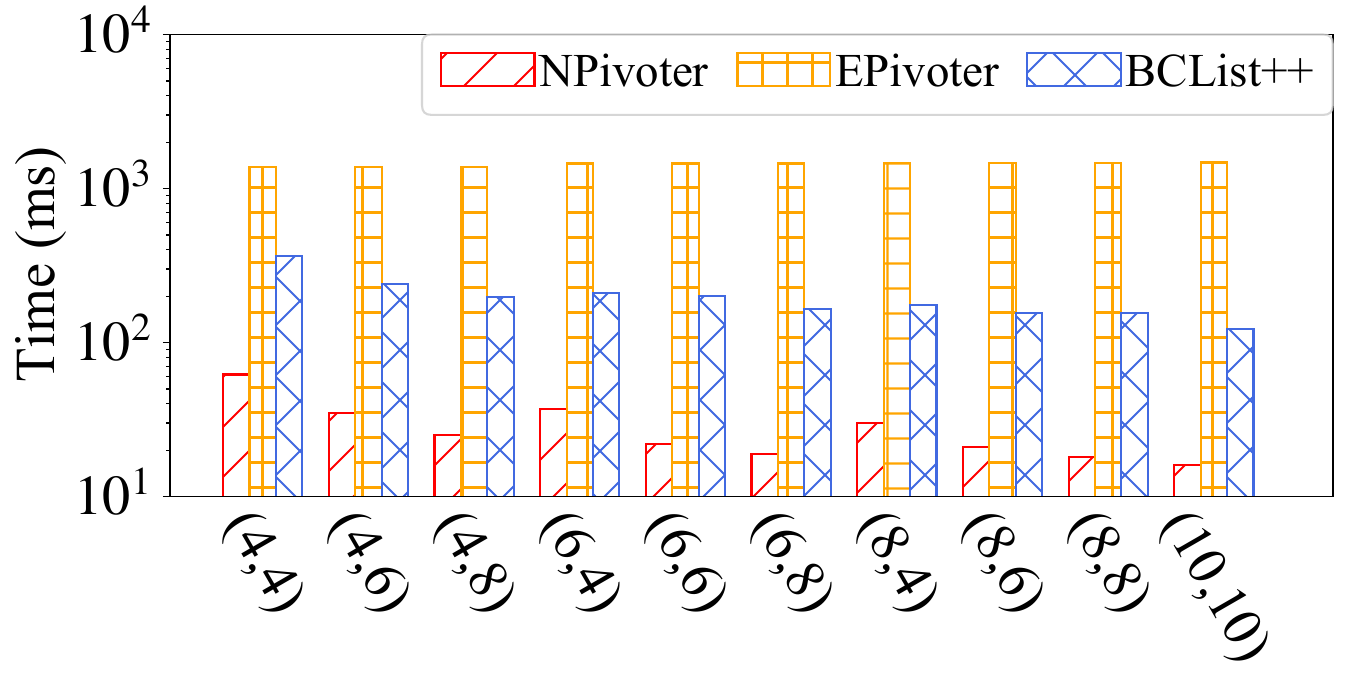}}
\subfigure[Git]{\label{sfig:time_git}\includegraphics[width=0.3\linewidth]{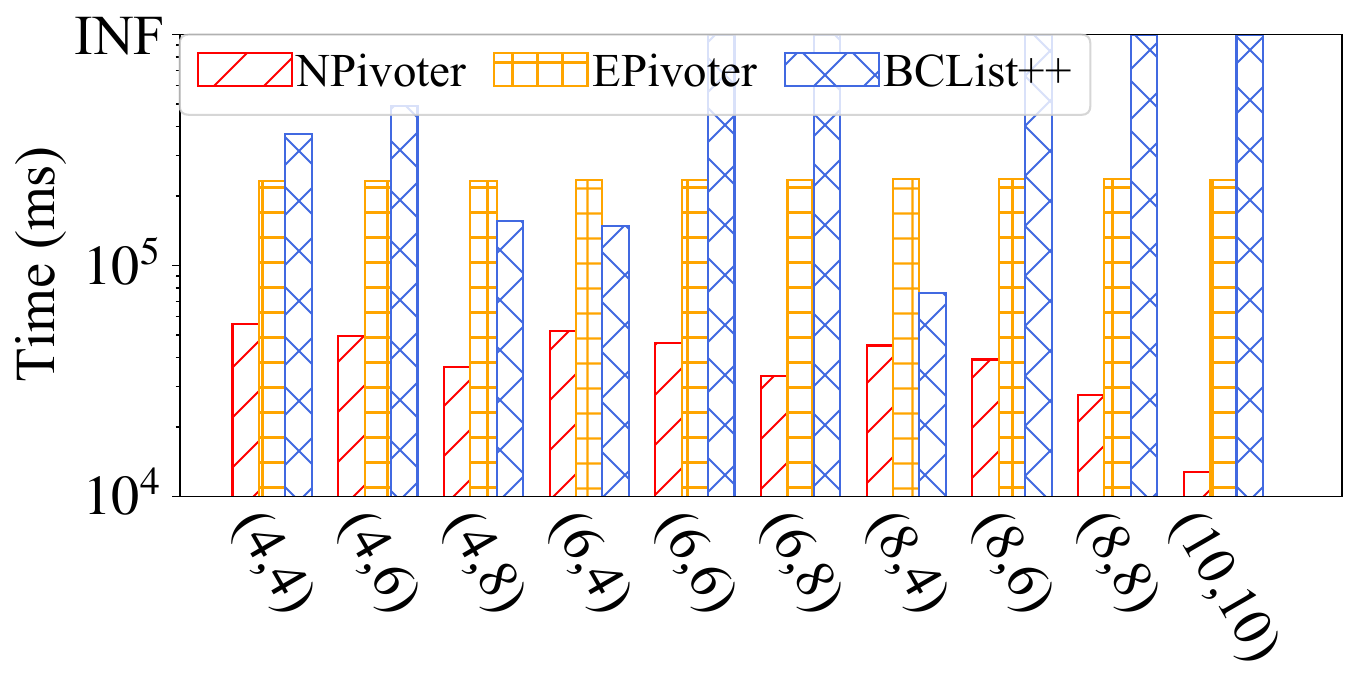}}
\subfigure[Cite]{\label{sfig:time_cite}\includegraphics[width=0.3\linewidth]{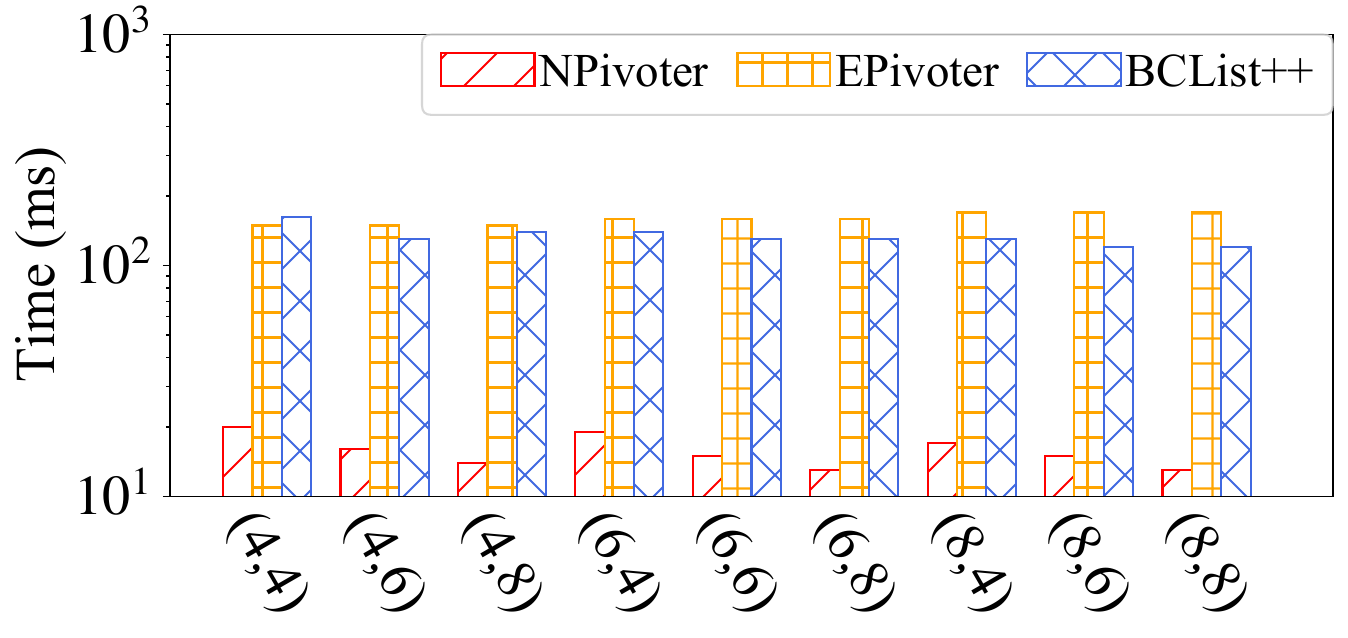}}
\subfigure[Stac]{\label{sfig:time_stackoverflow}\includegraphics[width=0.3\linewidth]{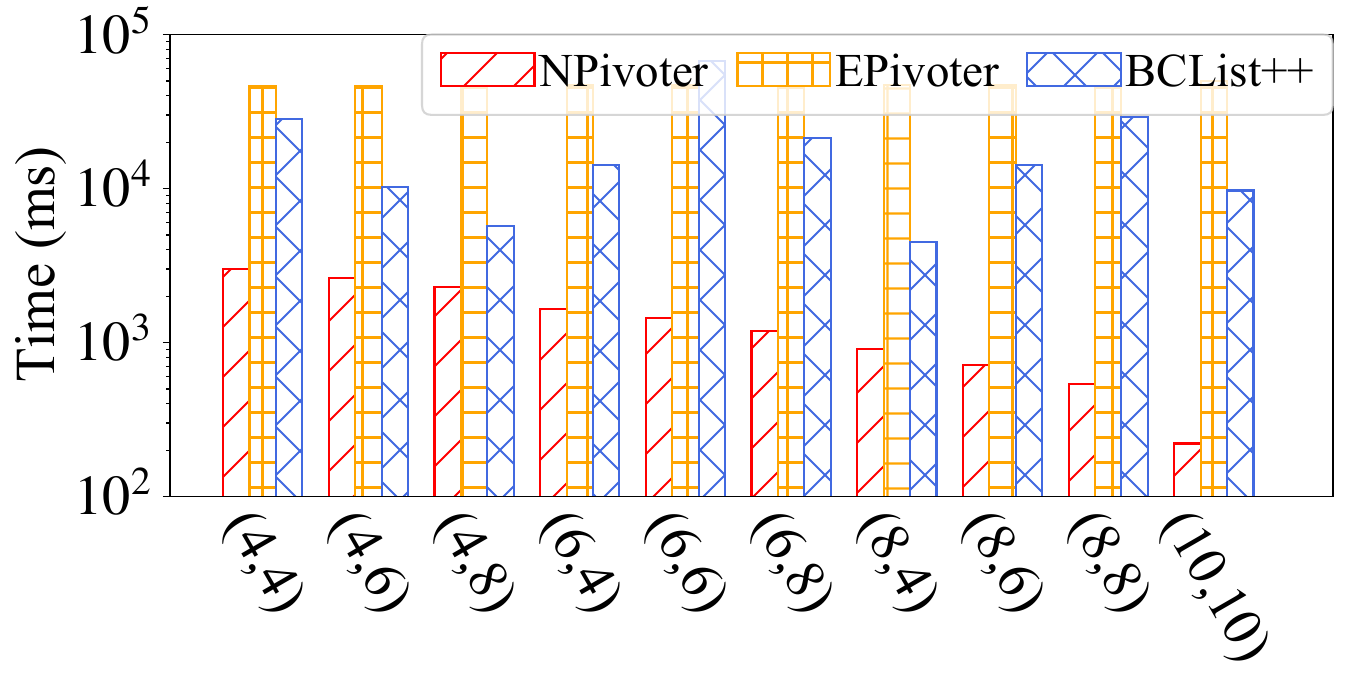}}
\subfigure[Twit]{\label{sfig:time_mumm}\includegraphics[width=0.3\linewidth]{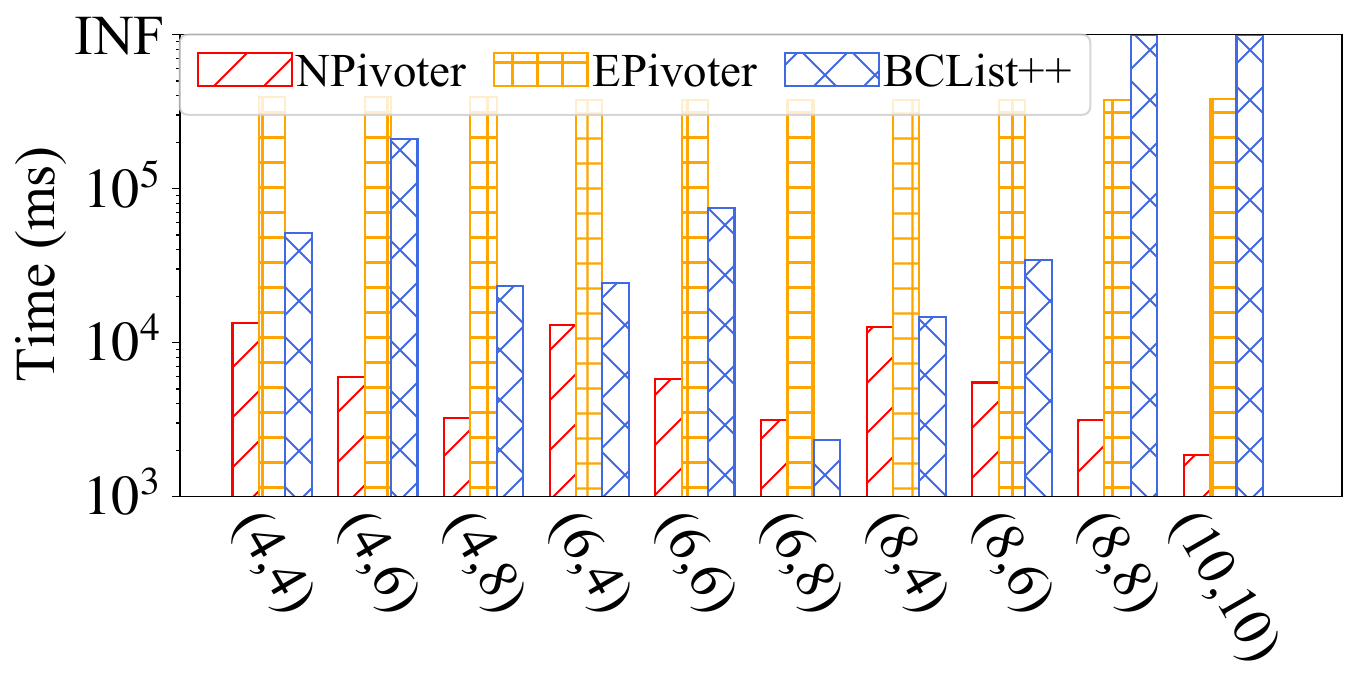}}
\subfigure[IMDB]{\label{sfig:time_imdb}\includegraphics[width=0.3\linewidth]{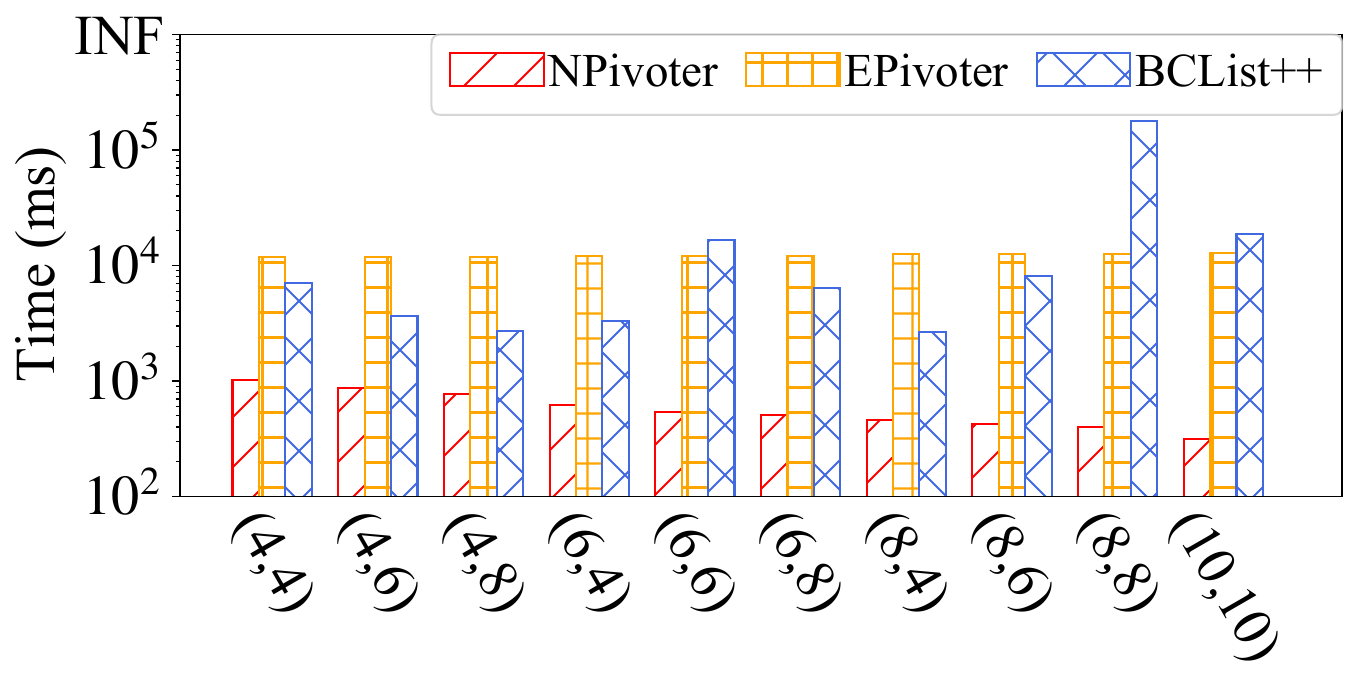}}
\subfigure[Genre]{\label{sfig:time_genre}\includegraphics[width=0.3\linewidth]{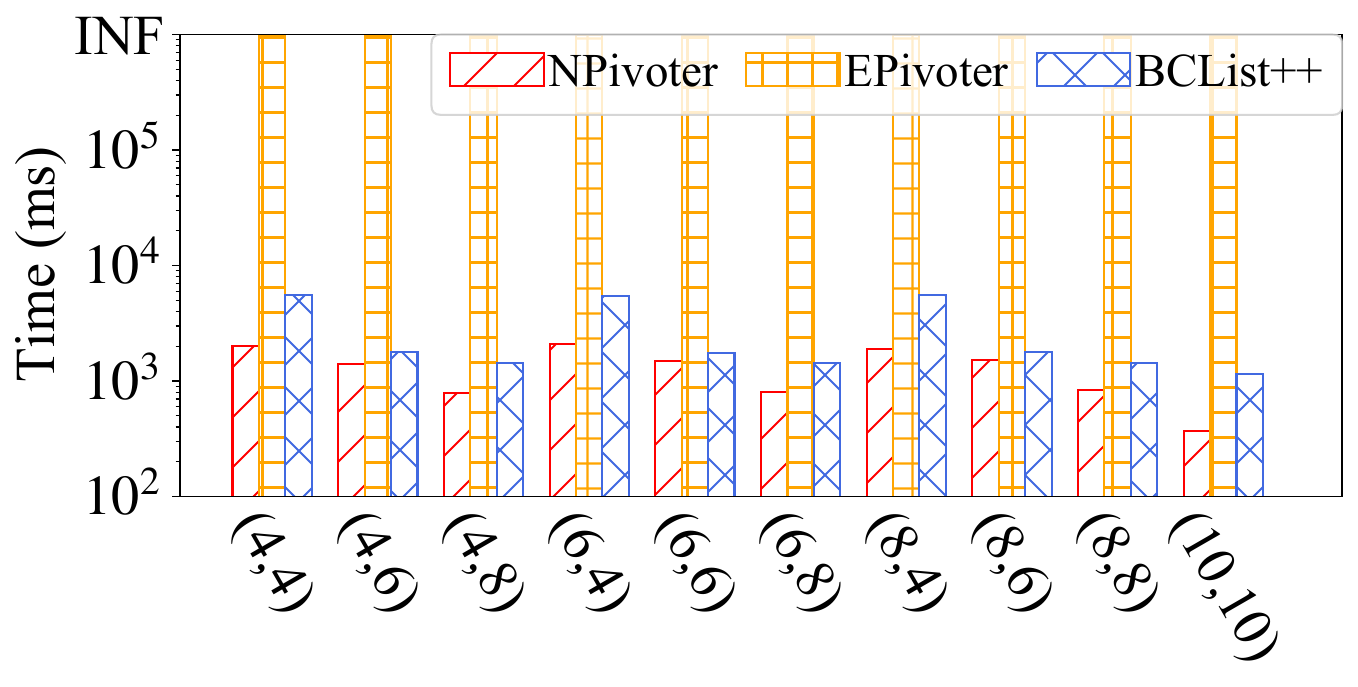}}
\subfigure[Act]{\label{sfig:time_actor2}\includegraphics[width=0.3\linewidth]{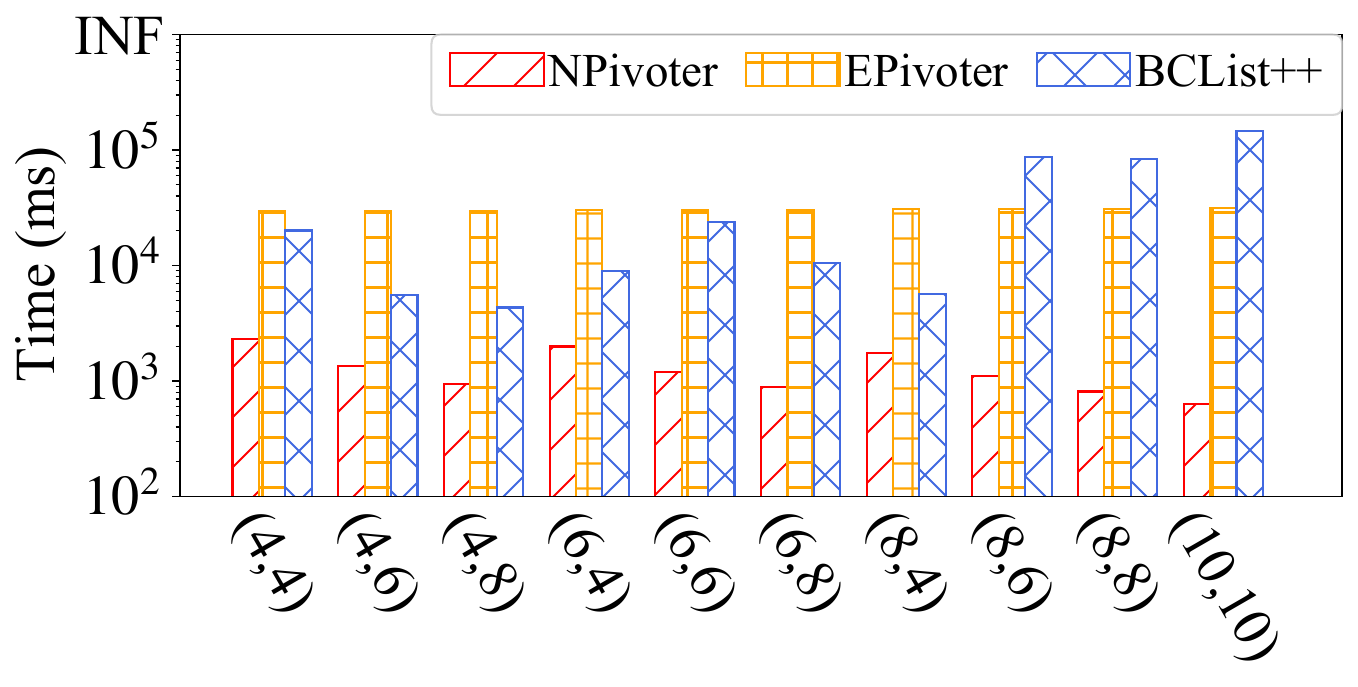}}
\subfigure[Ama]{\label{sfig:time_ama}\includegraphics[width=0.3\linewidth]{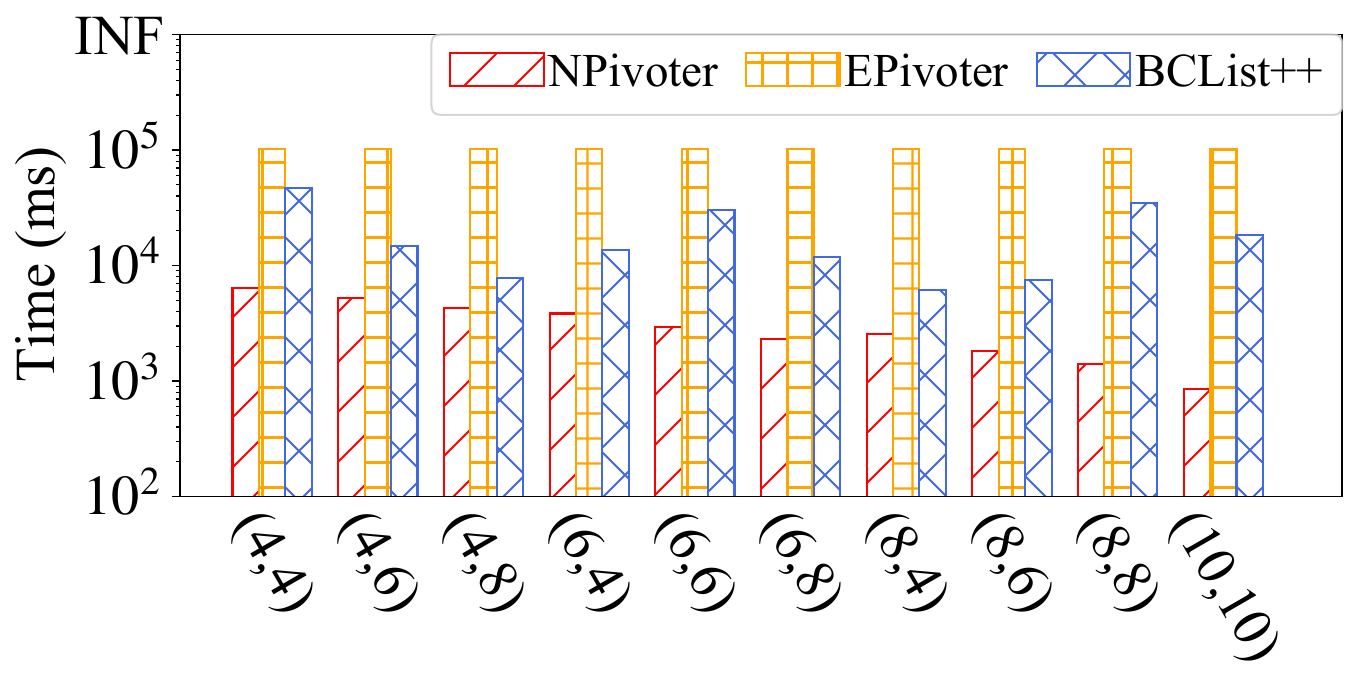}}
\subfigure[DBLP]{\label{sfig:time_dblp}\includegraphics[width=0.3\linewidth]{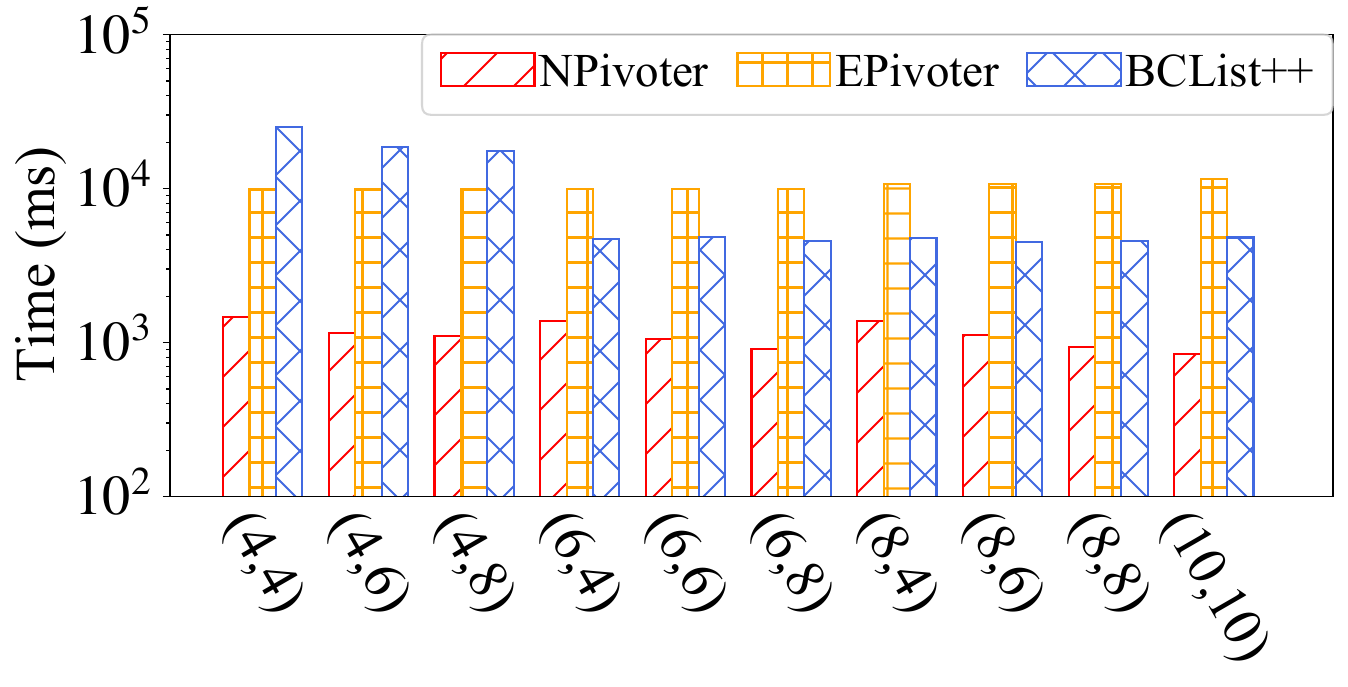}}
		\end{center}
%\vspace*{-0.2cm}
\caption{Running time of various algorithms}
\label{fig:time} 
%\vspace*{-0.1cm}
\end{figure*}

\begin{table}[t!]
	\small
	%	\scriptsize
	\centering
	\caption{The percentage of bicliques counted combinatorially}
	%	\vspace*{-0.3cm}
	\begin{tabular}{c|c|c| c|c}
		\toprule
		\multirow{2}{*}{\textbf{Datasets}} &
		$\bm{(4,4)}$  &$\bm{(6,6)}$  & $\bm{(8,8)}$ & $\bm{(10,10)}$ \\
		\cmidrule{2-5}
		\midrule
		You  & 74\%  & 96\%  & 100\%  & 100\% \\
		Kos  & 100\%  & 100\%  & 100\%  & 100\% \\
		Book  & 94\%  & 100\%  & 100\%  & 100\% \\
		Git  & 56\%  & 75\%  & 94\%  & 100\% \\
		Cite  & 100\%  & 100\%  & 100\%  & 100\% \\
		Stac  & 70\%  & 97\%  & 100\%  & 100\% \\
		Twit  & 91\%  & 96\%  & 94\%  & 100\% \\
		IMDB  & 95\%  & 100\%  & 100\%  & 100\% \\
		Genre  & 49\%  & 51\%  & 37\%  & 85\% \\
		Act  & 90\%  & 100\%  & 100\%  & 100\% \\
		Ama  & 73\%  & 79\%  & 93\%  & 100\% \\
		DBLP  & 100\%  & 100\%  & 100\%  & 100\% \\
		\bottomrule
	\end{tabular}
	%	\vspace*{-0.4cm}
	\label{tab:percentage}
\end{table}

\subsection{Empirical Results}
\stitle{Exp-1: Running Time of Various Algorithms.} Figure~\ref{fig:time} shows the running time of  \bc, \edgepivot, and \npivot with varying  $p$ and $q$ on 12 real-world networks. Across almost all cases, \npivot consistently demonstrates superior performance, significantly outperforming both \bc and \edgepivot. For instance, on the Twit network for $(10,10)$-biclique counting, \npivot is two orders of magnitude faster than \edgepivot. Similarly, on the Stac network for $(10,10)$-biclique counting, \npivot is two orders of magnitude faster than \bc. These results highlight the high efficiency of \npivot compared to other methods.

A notable and somewhat counter-intuitive observation is that the running time of \npivot decreases as the values of $p$ and $q$ increase. For example, on the Kos network, the running time decreases when moving from $(6,6)$ to $(6,12)$-biclique counting. Likewise, counting $(12,12)$-bicliques takes less time than counting $(9,9)$-bicliques or $(6,6)$-bicliques. This phenomenon is primarily due to two reasons: (1) larger values of $(p,q)$ enable more aggressive pruning of small-degree nodes according to the $(\alpha, \beta)$-core optimization (Section~\ref{sec:opt}), reducing the search space; (2) the impact of $p$ and $q$ on the search tree's size is minimal. Observe the \npc procedure in Algorithm~\ref{alg:npivot}, and we can find that $p$ and $q$ almost do not affect the search process except line~14. In effect, line~14 of Algorithm~\ref{alg:npivot} often terminates  due to empty edges instead of $H_U$ or $H_V$ reaching the maximum value, as shown in the following experiments. At last, this behavior demonstrates that \npivot is well-suited for efficiently counting $(p,q)$-bicliques, even for large values of $p$ and $q$.

\stitle{Exp-2: Effectiveness of the Proposed Node-pivot Technique.} Table~\ref{tab:percentage} presents the percentage of bicliques counting combinatorially across all datasets. A biclique is not counted combinatorially when it is handled in line 15 of Algorithm~\ref{alg:npivot} due to the condition \( |H_u| = p \) or \( |H_v| = q \) in line 14. As shown in Table~\ref{tab:percentage}, the majority of bicliques are counted in a combinatorial way. Furthermore, as the values of \(p\) and \(q\) increase, the percentage of combinatorially counted bicliques also increases. This is because larger values of \(p\) and \(q\) make it less likely to satisfy the condition \( |H_u| = p \) or \( |H_v| = q \) in line 14, demonstrating the growing effectiveness of the node-pivot technique with increasing biclique size. The results underscore the efficiency of the node-pivot method in reducing enumeration overhead and facilitating more efficient counting.

\begin{figure}[t]
	%	\vspace*{0.2cm}
	\begin{center}
		\subfigure[Ama]{\label{sfig:time_ama}\includegraphics[width=0.47\linewidth]{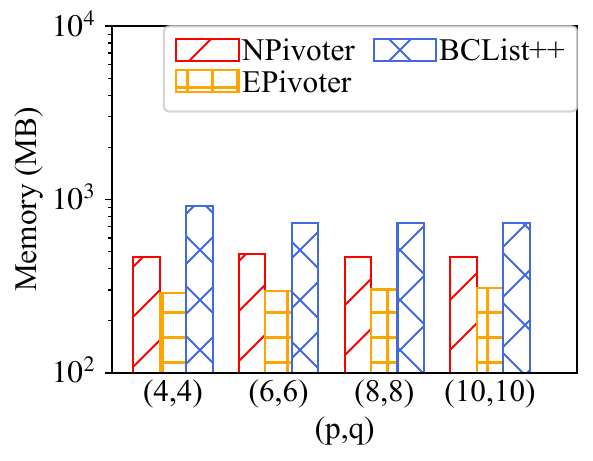}}
		\subfigure[DBLP]{\label{sfig:time_dblp}\includegraphics[width=0.47\linewidth]{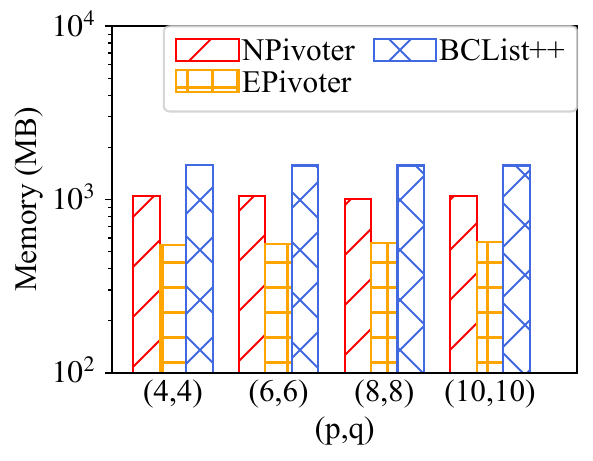}}
	\end{center}
	%\vspace*{-0.2cm}
	\caption{Memory overhead of all algorithms}
	\label{fig:mem}
	%\vspace*{-0.1cm}
\end{figure}

\stitle{Exp-3: Memory Overhead.} %Figure~\ref{fig:mem} shows the memory consumption of the evaluated algorithms. The results are computed by the "ps -aux" command of the Linux system. We can observe that both the three algorithms have similar order of memory cost. For example, on Ama, \npivot, \edgepivot and \bc costs $1048$ MB, $546$ MB, and $1578$ MB memory respectively. Among all the cases, \npivot costs no more than 2 times memory than \edgepivot. These results confirm the memory efficiency of our algorithms.
Figure~\ref{fig:mem} depicts the memory overheads of the evaluated algorithms: \npivot, \edgepivot, and \bc. The memory consumption was tracked using the ``ps -aux'' command on the Linux system. As can be seen, we can observe that the three algorithms exhibit similar magnitudes of memory usage across the datasets. For instance, on the Amazon (Ama) dataset, \npivot consumes 1,048 MB of memory, while \edgepivot and \bc consume 546 MB and 1,578 MB, respectively. Across all datasets, \npivot demonstrates memory efficiency, never using more than twice the memory of \edgepivot. In many cases, the difference in memory usage between the algorithms is even smaller. This indicates that although \npivot achieves significant performance improvements in terms of speed, it does not incur a significant increase in memory consumption. The results highlight that \npivot strikes a balanced trade-off between computational efficiency and memory overhead, making it well-suited for large-scale applications where both memory and time are critical factors.

\begin{figure}[t]
	%	\vspace*{0.2cm}
	\begin{center}
		\subfigure[Stac]{\label{sfig:time_stackoverflow}\includegraphics[width=0.94\linewidth]{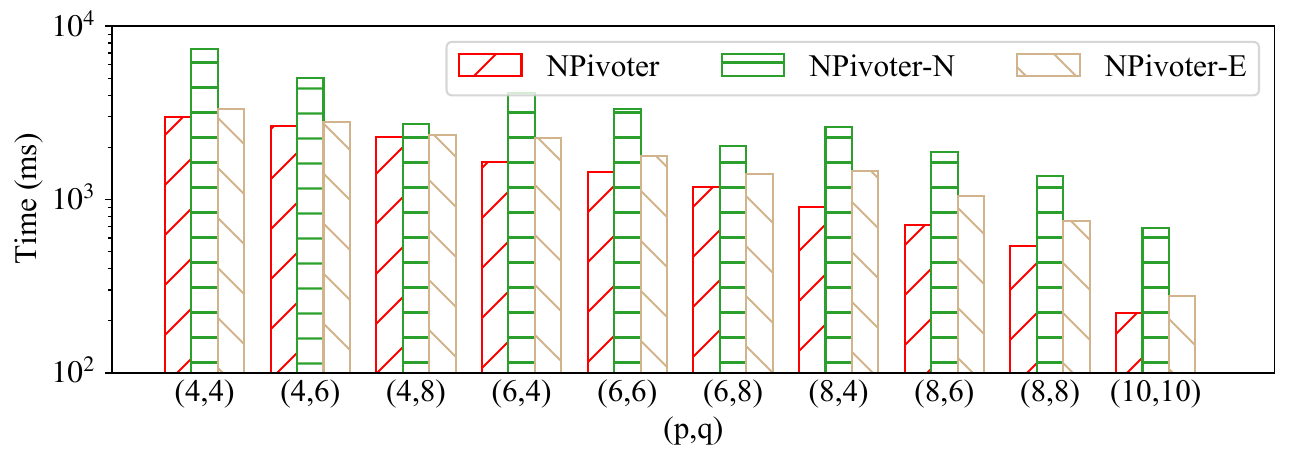}}
		
		\subfigure[Genre]{\label{sfig:time_genre}\includegraphics[width=0.94\linewidth]{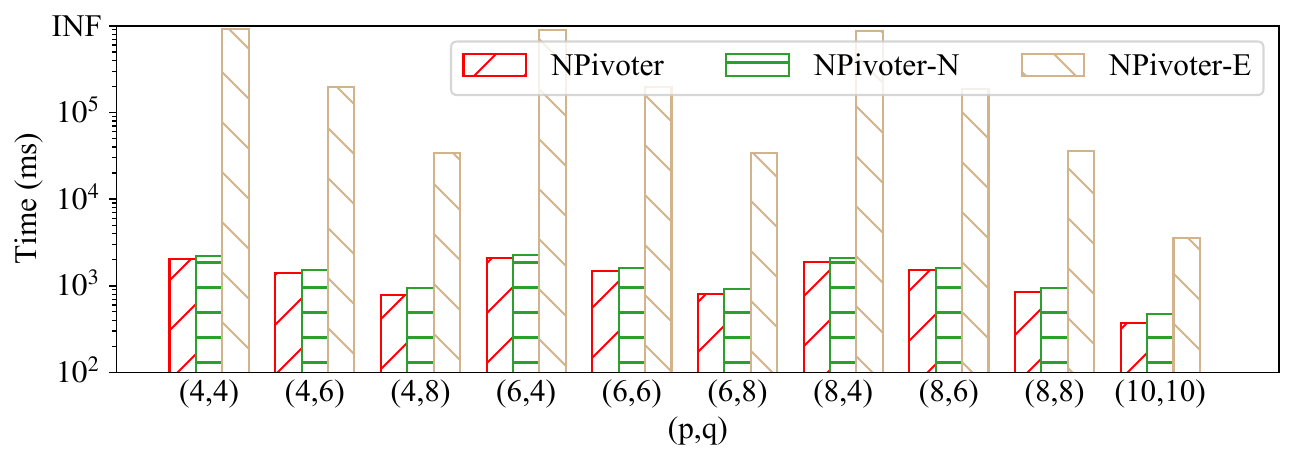}}
	\end{center}
	%\vspace*{-0.2cm}
	\caption{Effect of the cost estimator}
	\label{fig:effectOfCost}
	%\vspace*{-0.1cm}
\end{figure}

\begin{table}[t!]
	\small
	%	\scriptsize
	\centering
	\caption{The time of building cost estimation index (ms) and the percentage of the nodes using the node-split strategy}
	%	\vspace*{-0.3cm}
	\begin{tabular}{c|c|c|c|c}
		\toprule
		\multirow{2}{*}{\textbf{Datasets}} &
		\multicolumn{4}{c}{$\bm{(x, y)}$ }  \\
		\cmidrule{2-5}
		& $\bm{(4,4)}$ & $\bm{(6,6)}$  & $\bm{(8,8)}$  & $\bm{(10,10)}$ \\
		\midrule
		% You & 637 / 3.3\% & 563 / 0.7\% & 500 / 0.2\% & 452 / 0.1\% \\
		% Kos & 11411 / 6.3\% & 11076 / 5.3\% & 10838 / 5.2\% & 10659 / 5.0\% \\
		% Book & 649 / 0.3\% & 546 / 0.0\% & 476 / 0.0\% & 423 / 0.0\% \\
		% Git & 8856 / 6.4\% & 8099 / 2.0\% & 7357 / 0.6\% & 6702 / 0.1\% \\
		% Cite & 45 / 1.1\% & 38 / 0.0\% & 34 / 0.0\% & 29 / 0.0\% \\
		% Stac & 12866 / 2.2\% & 11690 / 0.4\% & 10748 / 0.1\% & 9827 / 0.0\% \\
		% Twit & 233601 / 0.0\% & 191136 / 3.1\% & 155206 / 1.0\% & 123646 / 0.3\% \\
		% IMDB & 11577 / 3.8\% & 11049 / 0.6\% & 10283 / 0.1\% & 9319 / 0.0\% \\
		% Genre & 204 / 80.0\% & 138 / 66.7\% & 113 / 53.3\% & 68 / 40.0\% \\
		% Act & 21956 / 3.5\% & 20246 / 0.5\% & 18085 / 0.1\% & 15548 / 0.0\% \\
		% Ama & 52870 / 0.5\% & 39982 / 0.1\% & 32887 / 0.0\% & 27451 / 0.0\% \\
		% DBLP & 1789 / 1.4\% & 1638 / 0.0\% & 1383 / 0.0\% & 1343 / 0.0\% \\

  You & 28 / 3.3\% & 23 / 0.7\% & 20 / 0.2\% & 17 / 0.1\% \\
Kos & 217 / 6.3\% & 213 / 5.3\% & 203 / 5.2\% & 193 / 5.0\% \\
Book & 24 / 0.3\% & 20 / 0.0\% & 18 / 0.0\% & 17 / 0.0\% \\
Git & 148 / 6.4\% & 123 / 2.0\% & 107 / 0.6\% & 95 / 0.1\% \\
Cite & 17 / 1.1\% & 14 / 0.0\% & 13 / 0.0\% & 12 / 0.0\% \\
Stac & 209 / 2.2\% & 180 / 0.4\% & 164 / 0.1\% & 149 / 0.0\% \\
Twit & 2618 / 13.9\% & 2065 / 3.1\% & 1681 / 1.0\% & 1310 / 0.3\% \\
IMDB & 287 / 3.8\% & 259 / 0.6\% & 240 / 0.1\% & 220 / 0.0\% \\
Genre & 100 / 80.0\% & 61 / 66.7\% & 43 / 53.3\% & 27 / 40.0\% \\
Act & 503 / 3.5\% & 451 / 0.5\% & 404 / 0.1\% & 360 / 0.0\% \\
Ama & 1058 / 0.5\% & 797 / 0.1\% & 655 / 0.0\% & 564 / 0.0\% \\
DBLP & 761 / 1.4\% & 642 / 0.0\% & 555 / 0.0\% & 510 / 0.0\% \\
		\bottomrule
	\end{tabular}
	%	\vspace*{-0.4cm}
	\label{tab:costOfIdx}
\end{table}

%\stitle{Exp-3: Performance of the cost estimator.} 
%Figure~\ref{fig:effectOfCost} shows the effect of the cost estimator. \npivot-N is the \npivot framework that only use the node-split strategy. \npivot-E is the \npivot framework that only use the edge-split strategy. We choose two representative networks. The experiment results are similar on the other networks. On the Stac network, \npivot-E is more efficient than \npivot-N. Differently, on the Genre network, \npivot-N is more efficient than \npivot-E. However, on both the two networks, \npivot outperforms \npivot-N and \npivot-E. The results prove the effectiveness of our cost estimator.

%Table~\ref{tab:costOfIdx} shows the time of building the cost estimator index and the percentage of node-split. (1) We can find that the time of building the cost estimator index is not  sensitive to the parameter $(x, y)$, where with the increase of $(x,y)$, the building time varies a little. For example, on the You network, $(10,10)$ is only $1.4\times$ faster than $(4,4)$. (2)  The real-world networks usually submit to the power law, which makes the `dense part' of the networks usually account for a small proportion of nodes \cite{}. Except the Genre network, the percentage of nodes using the node-split strategy is very small. This is because the `dense part' nodes tend to use the node-split strategy and  the `sparse part' nodes tend to use the edge-split strategy. The Genre network differs because it only has $15$ nodes at one side (Table~\ref{tab:datasets}) and thus most of them are in `dense part'.

\stitle{Exp-4: Effect of the Cost Estimator.}  
Figure~\ref{fig:effectOfCost} shows the impact of the cost estimator on the performance of \npivot. To analyze this, we compare \npivot with two versions that exclusively employ either the node-split or the edge-split strategy: \npivot-N (node-split only) and \npivot-E (edge-split only). We conducted experiments on two representative datasets, the Stac and Genre datasets, though the results were consistent across other networks as well.  Specifically, on the Stac dataset, \npivot-E—using only the edge-split strategy—proves to be more efficient than \npivot-N, which exclusively applies the node-split strategy. This suggests that in sparser networks like Stac, edge-split is generally more effective due to the structure of the graph. Conversely, on the Genre network, \npivot-N performs significantly better than \npivot-E. This is due to the graph’s smaller size and denser connections, where node-split is better suited to handle dense subgraphs. Despite the differences in performance between \npivot-N and \npivot-E across these graphs, \npivot, which uses the cost estimator to dynamically choose between the two strategies, consistently outperforms both. This clearly demonstrates the effectiveness of our cost estimator, which intelligently selects the more efficient split strategy based on the graph structure, leading to improved overall performance.

\stitle{Exp-5: The Performance of Cost Estimator Index.}  
Table~\ref{tab:costOfIdx} shows two key aspects: the time required to build the cost estimator index and the proportion of nodes that employ the node-split strategy.  The results show that building the index consumes quite less time than counting $(p,q)$-biclique, ensuring that building the index is not a major bottleneck in the algorithm. For instance, on the You dataset, counting takes 1989 ms, while building the index only takes 20 ms. First, the results indicate that the time taken to build the index is not significantly affected by the parameter values $(x, y)$. As $(x, y)$ increases, the variation in index construction time is minimal. For example, on the You dataset, the index construction for $(10,10)$ is only 1.6 times faster than that for $(4,4)$. This indicates that the index construction time remains stable across different parameter settings. 

Second, the results show that in most real-world datasets, a relatively small proportion of nodes use the node-split strategy. This is because real-world networks typically follow a power-law distribution, where a small subset of nodes (the ``dense part'') have a disproportionately high degree, while the majority of nodes (the ``sparse part'') have fewer connections. Consequently, the denser nodes tend to benefit from the node-split strategy, whereas the sparser nodes are better suited for the edge-split strategy. The exception is the Genre network, where most of the nodes employ the node-split strategy. This is due to the fact that Genre has only 15 nodes on one side of the bipartite graph, most of which are densely connected (as shown in Table~\ref{tab:datasets}). Therefore, in this case, the node-split strategy dominates. 

These findings reinforce the practical utility of the cost estimator, which adapts well to varying network structures by efficiently deciding whether to use node-split or edge-split based on the local graph characteristics. This flexibility contributes to the strong performance of \npivot across diverse real-world datasets.

\begin{table}[t!]
	\small
	%	\scriptsize
	\centering
	\caption{The time cost of local counting for each node (ms)}
	%	\vspace*{-0.3cm}
	\begin{tabular}{c|cc|cc| cc}
		\toprule
		\multirow{2}{*}{\textbf{Datasets}} &
		\multicolumn{2}{c|}{$\bm{(4,4)}$ } & \multicolumn{2}{c|}{$\bm{(6,6)}$ } & \multicolumn{2}{c}{$\bm{(8,8)}$ } \\
		\cmidrule{2-7}
		& \textbf{Local} & \textbf{Global}  & \textbf{Local} & \textbf{Global}  & \textbf{Local} & \textbf{Global} \\
		\midrule
		You & 2195 & 1989 & 1448 & 1325 & 745 & 704 \\
		Kos & 378266 & 349997 & 324288 & 303924 & 198834 & 194317 \\
		Book & 66 & 62 & 24 & 22 & 20 & 18 \\
		Git & 61772 & 55600 & 50100 & 46091 & 28096 & 27441 \\
		Cite & 22 & 20 & 16 & 15 & 14 & 13 \\
		Stac & 3262 & 2994 & 1524 & 1437 & 557 & 539 \\
		Twit & 14591 & 13478 & 6395 & 5821 & 3516 & 3138 \\
		IMDB & 1122 & 1025 & 603 & 541 & 440 & 401 \\
		Genre & 2240 & 2013 & 1641 & 1493 & 960 & 841 \\
		Act & 2551 & 2310 & 1330 & 1206 & 914 & 812 \\
		Ama & 6975 & 6418 & 3126 & 2948 & 1532 & 1410 \\
		DBLP & 1553 & 1465 & 1265 & 1050 & 1113 & 930 \\
		\bottomrule
	\end{tabular}
	%	\vspace*{-0.4cm}
	\label{tab:local}
\end{table}

%\stitle{Exp-4: Local counting.} The time of local counting for each node is shown in Table~\ref{tab:local}. The columns `Local' are time of local counting $(p,q)$-biclique for each node. The columns `Global' are time of counting $(p,q)$-biclique in the whole network. As shown in Table~\ref{tab:local}, local counting only takes at most  $1.2\times $ more time than global counting. The results demonstrate the high efficiency of our \npivot on local counting.

\stitle{Exp-6: Local Counting.}  
Table~\ref{tab:local} reports the time taken for local counting of $(p,q)$-bicliques for each node, alongside the time required for global counting in the entire network. The columns labeled `Local' correspond to the time spent in performing local counting, i.e., counting $(p,q)$-bicliques that involve each individual node. In contrast, the `Global' columns show the time taken for counting all $(p,q)$-bicliques across the entire graph. As demonstrated in Table~\ref{tab:local}, local counting proves to be highly efficient, as it requires only marginally more time than global counting—at most $1.2\times$ longer. This is an important observation, as one might expect local counting to be significantly slower, given that it isolates the counting process for each node individually. However, the results indicate that our \npivot algorithm handles local counting with minimal overhead, even when compared to the time needed for global counting over the entire graph.

The high efficiency of local counting is crucial in applications where the focus is on understanding the role or influence of specific nodes within the graph. For example, in social networks, it may be important to locally analyze the influence of certain users or groups. In recommendation systems, local counting allows for examining how particular items or users contribute to specific biclique patterns. Similarly, in citation networks, local counting helps analyze the impact of individual authors or papers on collaboration patterns. Deleted in the full version.

%The results highlight the flexibility and robustness of \npivot in handling both global and local counting tasks. By maintaining near-optimal performance for local counting, \npivot demonstrates its practicality for use in scenarios that demand node-specific analysis without incurring significant additional computational costs. This capability further solidifies \npivot as a versatile and efficient tool for biclique counting in large-scale real-world networks.

\begin{table}[t!]
	\small
	%\scriptsize
	\centering
	%	\vspace*{-0.1cm}
	\caption{Comparing  the running time of  only counting $(4,4)$-biclique to counting all $(p,q)$-bicliques with $p\in [4, 10], q\in [4, 10]$ simultaneously (ms)} %\vspace*{-0.3cm}
	\begin{tabular}{c| cc|  cc}
		\toprule
		\multirow{2}{*}{\textbf{Datasets}} 
		&   \multicolumn{2}{c|}{\npivot} &
		\multicolumn{2}{c}{\edgepivot} \\
		\cline{2-5}
		&   ${(4,4)}$ &   all 
		&   ${(4,4)}$ &  all \\
		\midrule
		
		You & 298 & 2244 & 13939 & 16338 \\
		Kos & 104967 & 403350 & $>10^6$ & $>10^6$ \\
		Book & 16 & 64 & 1437 & $>10^6$ \\
		Git & 12780 & 64558 & 447438 & 495141 \\
		Cite & 12 & 21 & 180 & $>10^6$ \\
		Stac & 221 & 3206 & 51735 & $>10^6$ \\
		Twit & 1851 & 14157 & $>10^6$ & $>10^6$ \\
		IMDB & 313 & 1110 & 12361 & 16180 \\
		Genre & 369 & 2501 & $>10^6$ & $>10^6$ \\
		Act & 631 & 2467 & 30917 & 38346 \\
		Ama & 854 & 6801 & 103109 & 115257 \\
		DBLP & 838 & 1813 & 10700 & 17848 \\
		\bottomrule
	\end{tabular}
	%	\vspace{-0.25cm}
	\label{tab:range}
\end{table}

\stitle{Exp-7: Range Counting.}  
Table~\ref{tab:range} provides a comparative analysis of the time required for two different counting scenarios: counting only $(4,4)$-bicliques versus counting all $(p,q)$-bicliques where $p \in [4, 10]$ and $q \in [4, 10]$ simultaneously. In  Table~\ref{tab:range} , we observe that range counting with \npivot takes an order of magnitude longer than counting a single $(4,4)$-biclique. For instance, on the You network, \npivot requires $298$ ms to count $(4,4)$-bicliques but takes $2244$ ms to perform range counting. This increase in computation time is expected due to the additional complexity involved in managing multiple parameters across a range of values, which inherently requires more extensive processing. Despite this, \npivot's performance in range counting remains significantly superior to that of the state-of-the-art algorithm \edgepivot. On the IMDB network, for example, \npivot is $14.6$ times faster than \edgepivot for range counting tasks. This substantial efficiency advantage highlights \npivot's effectiveness and scalability in handling complex counting scenarios where multiple $(p,q)$-biclique sizes need to be evaluated simultaneously.

%\stitle{Exp-6: Scalability.} We generate random graphs to show the scalability of our \npivot. We fix the number of edges to $|E|=8000$. Then, for each $a\in [20,70], b\in [20,70]$, generate $20$ random graphs with $\overline{d}_U  = a$ and $\overline{d}_V=b$. Each color grid in Figure~\ref{fig:scalability} is the average running time on the 20 random networks. As shown in Figure~\ref{fig:scalability}, the running time stays smaller than $1000$ ms when $\overline{d}_U<60$ or   $\overline{d}_V < 60$. The experiment results indicate that our \npivot is scalable for sparse networks.
\begin{figure}[t]
	%	\vspace*{0.2cm}
	\begin{center}
		\subfigure[sampling edges]{\label{sfig:edge}\includegraphics[width=0.47\linewidth]{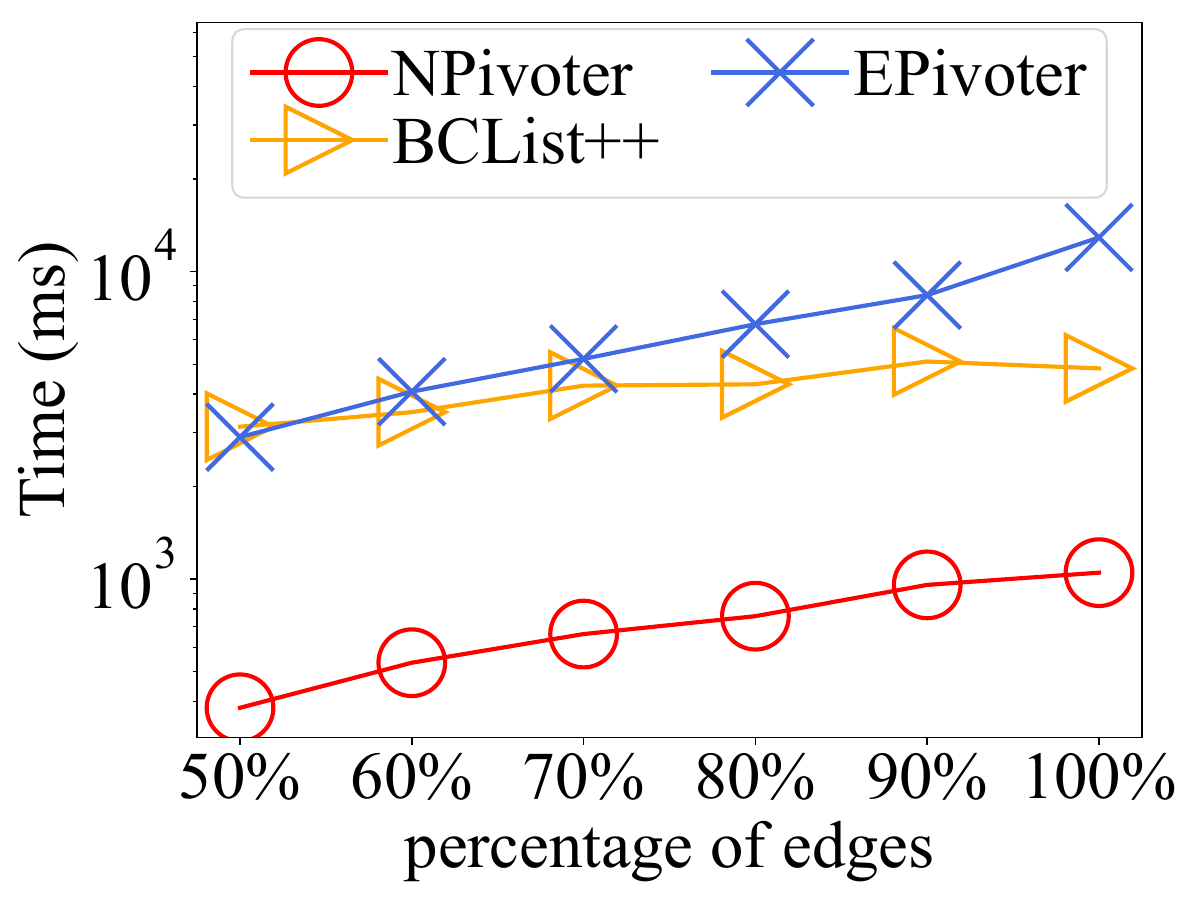}}
		\subfigure[sampling nodes]{\label{sfig:node}\includegraphics[width=0.47\linewidth]{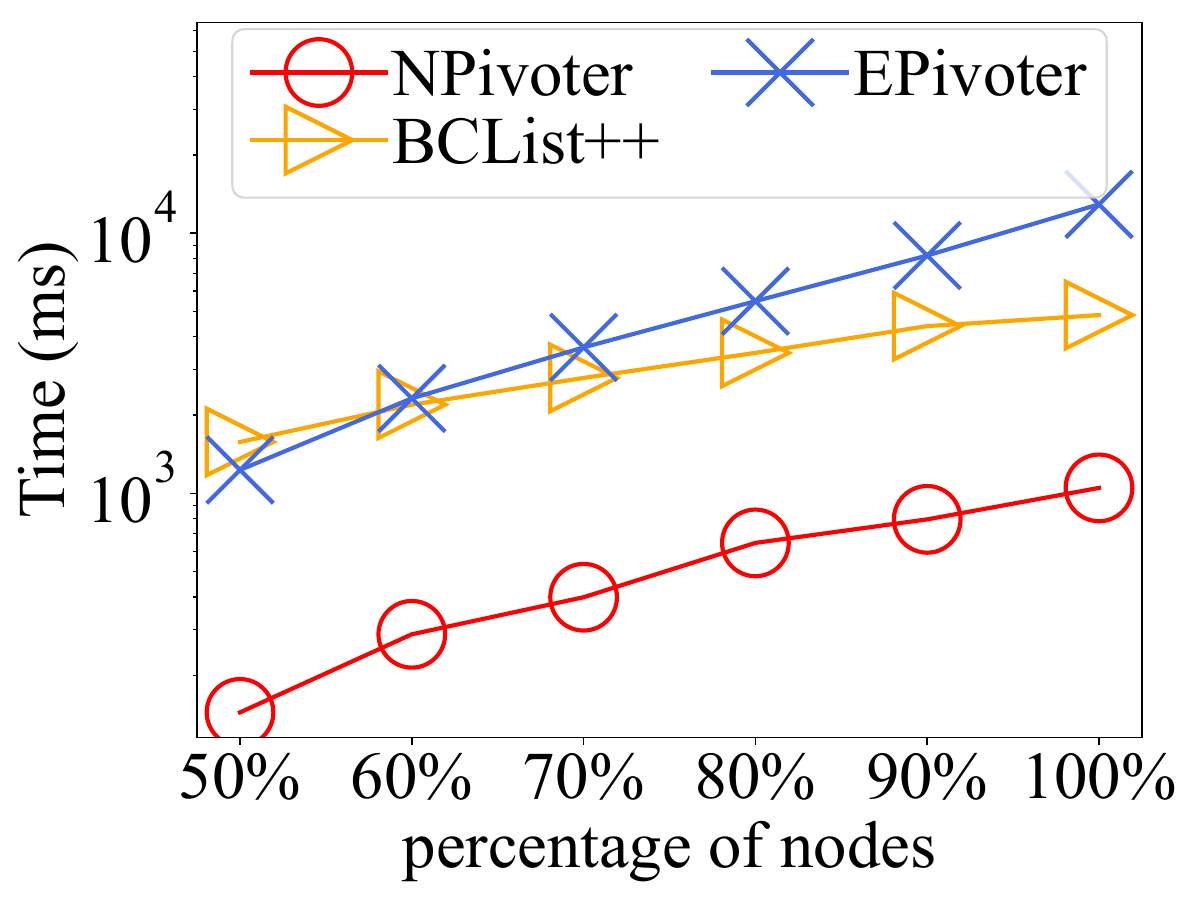}}
	\end{center}
	%\vspace*{-0.2cm}
	\caption{\textcolor{black}{Scalability test}}
	\label{fig:percentage_scalability}
	%\vspace*{-0.1cm}
\end{figure}

\stitle{Exp-8: Scalability Test.} In Figure~\ref{fig:percentage_scalability}, we evaluate the scalability of \npivot by adjusting the size of datasets. We sample either the edges or nodes, varying the percentage from $50\%$ to $100\%$. In this experiment, we use the DBLP dataset, which has the highest number of edges among the tested datasets. We set $p=6, q=6$. The results obtained on the other datasets and with other $p,q$ values show similar trends. From Figure~\ref{fig:percentage_scalability}, we can see that all three algorithms—\npivot, \bc, and \edgepivot— exhibit a good scalability with respect to graph size.   Furthermore, \npivot outperforms both \bc and \edgepivot significantly, showcasing its superior efficiency and capacity to handle large-scale datasets.

\begin{table}[t!]
	\small
	%\scriptsize
	\centering
	%	\vspace*{-0.1cm}
	\caption{\textcolor{black}{The effect of further optimizations (ms); $O_1$: $(\alpha,\beta)$-core	based graph reduction; $O_2$: core value-based node ranking, $O_3$: early termination, $O_4$: maintenance of non-neighbor count}} %\vspace*{-0.3cm}
	\begin{tabular}{c| c |c |c |c}
		\toprule
		Datasets & You & Kos & DBLP & Ama \\
		\midrule
		
		No Optimization		& 13181 & 1138038 & 92856 & 4622 \\
		$O_1$	& 10851 & 1130167 & 15823 & 1971 \\
		$O_1 + O_2$		& 1793 & 392365 & 4195 & 1157 \\
		$O_1 + O_2 + O_3$	& 1341 & 319504 & 3043 & 1127 \\
		$O_1 + O_2 + O_3 + O_4$	& 1325 & 303924 & 2948 & 1050 \\
		
		\bottomrule
	\end{tabular}
	%	\vspace{-0.25cm}
	\label{tab:opt}
\end{table}

\stitle{Exp-9: The Effect of the Optimization Tricks.}
\textcolor{black}{
Table~\ref{tab:opt} reports the impact of four key optimizations  (presented in Section~\ref{sec:opt}) applied to the \npivot algorithm on several datasets: You, Kos, DBLP, and Ama. The results on the other datasets are similar. The optimizations include: $(\alpha, \beta)$-core based graph reduction ($O_1$), core value-based node ranking ($O_2$), early termination ($O_3$), and non-neighbor count maintenance ($O_4$). With each additional optimization, the running time consistently decreases. For instance, on the Ama dataset, the running time reduces by 57.3\% when applying $O_1$, followed by a further 41.6\% decrease when adding $O_2$, a 2.6\% reduction with $O_3$, and an additional 6.8\% improvement with $O_4$.   The full set of optimizations ($O_1 + O_2 + O_3 + O_4$) yields the best performance, with the runtime for all datasets minimized, indicating the effectiveness of these optimizations in enhancing algorithm efficiency.
}

\begin{figure}[t!]
	%	\vspace*{0.2cm}
	\begin{center}
		\includegraphics[width=0.9\linewidth]{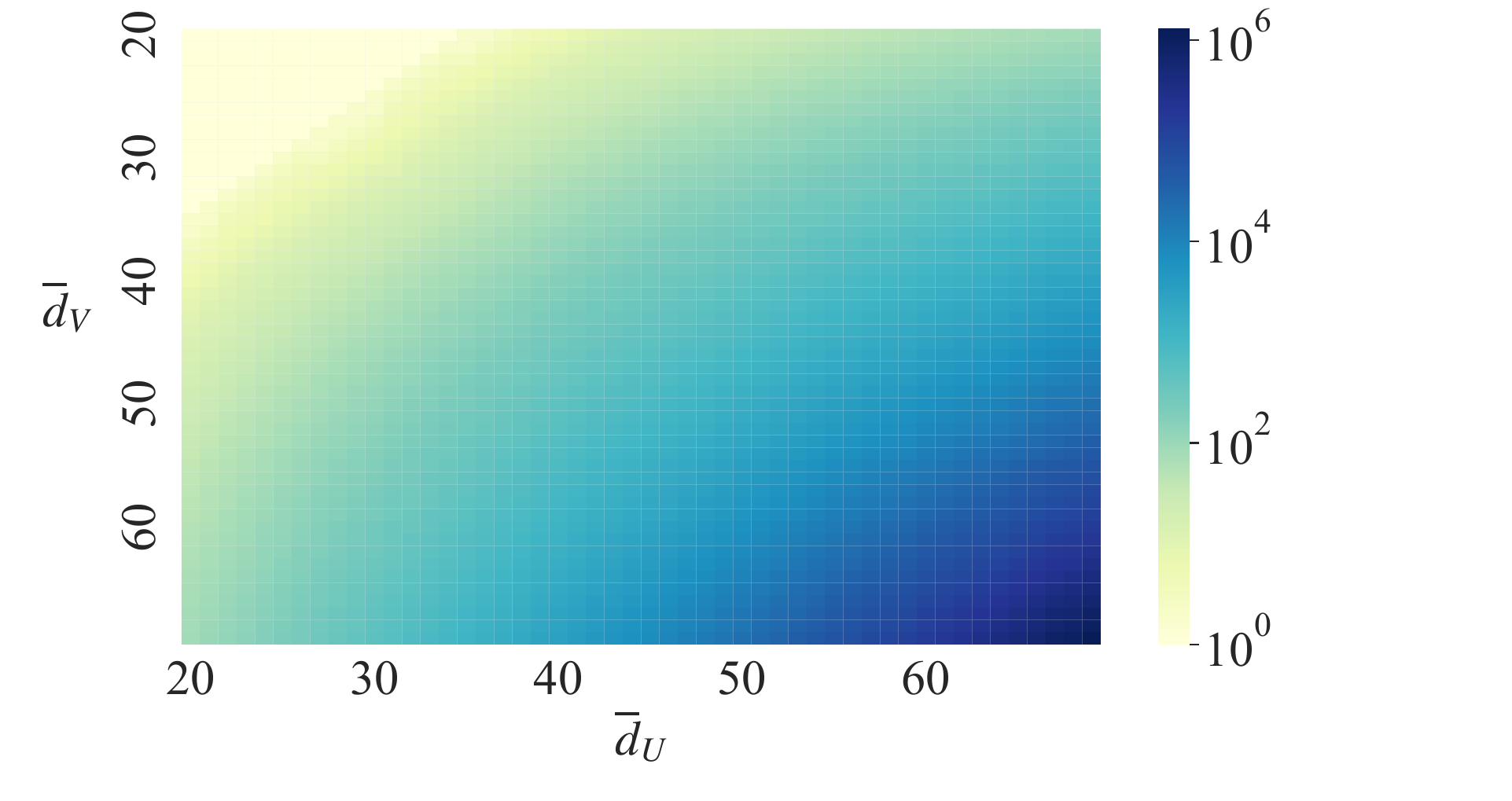}
	\end{center}
	%\vspace*{-0.2cm}
	\caption{The impact of graph density (ms)}
	\label{fig:scalability}
	%\vspace*{-0.1cm}
\end{figure}
\stitle{Exp-10: The Impact of Graph Density.}  
To evaluate the scalability of \npivot, we generate a set of random graphs under varying density. Specifically, we fix the number of edges at $|E| = 8000$ and generate random bipartite graphs with varying average degrees. For each combination of $\overline{d}_U \in [20,70]$ and $\overline{d}_V \in [20,70]$, we create 20 random graphs, leading to a total of 50,000 experiments. The goal is to assess how \npivot behaves as the density of the graph changes.

Each color grid in Figure~\ref{fig:scalability} represents the average running time for the 20 random graphs corresponding to a particular combination of $\overline{d}_U$ and $\overline{d}_V$.  Notably, the running time remains under 1000 ms when $\overline{d}_U < 60$ or $\overline{d}_V < 60$. This pattern indicates that \npivot scales well with the sparsity of the graph, making it a highly effective solution for real-world networks that tend to exhibit a power-law distribution with a majority of nodes having low degrees. As $\overline{d}_U$ and $\overline{d}_V$ increase, the computational complexity naturally rises due to the increase in possible connections between the two sets. However, the results show that \npivot handles this increase in density in a controlled manner. Even at the higher end of the degree spectrum, where both $\overline{d}_U$ and $\overline{d}_V$ approach 70, the algorithm still performs efficiently, although the running time does begin to rise sharply. These findings show the high scalability of \npivot, particularly for handling sparse networks. Since real-world networks are often sparse, these results further validate the practical applicability of \npivot in real-world scenarios. 

\section{related work}
\stitle{Subgraph Counting.} 
Biclique counting is a specific case of the broader problem of subgraph counting, which has been extensively studied in graph theory and its applications. The subgraph counting problem can be categorized based on several dimensions: Exact subgraph counting aims to compute the precise number of occurrences of a specific subgraph within a graph \cite{OrbitsCounting, 17WWWmotif, 20wsdmmotif, graft2014}, while approximate methods trade accuracy for scalability in large networks \cite{pathsampling2015, pathsampling2018, colorcodingTkdd, 09KDDtriangle}. Learning-based approaches also exist as approximate methods \cite{DBLP:conf/icde/HouZT24, DBLP:journals/vldb/ZhaoYLZR23}. Exact counting problems are closely related to enumeration problems, and  algorithms that performs exact counting can often be used for enumeration. Global counting involves determining the number of subgraphs across the entire graph \cite{OrbitsCounting, 17WWWmotif}, while local counting focuses on computing subgraph counts for specific nodes or edges \cite{DBLP:journals/pvldb/LiY24}. In this work, biclique counting falls under both categories, as applications may require global counts or local counts per node, as seen in social networks or graph neural networks. While some algorithms are designed for counting specific subgraph patterns like cliques, cycles, or bicliques \cite{DBLP:journals/siamcomp/FockeR24, DBLP:conf/stoc/DoringMW24,edgepivot,BClistYangPZ21,PIVOTER}, others focus on counting a wide variety of subgraph types \cite{ColorCoding,DBLP:journals/vldb/ZhaoYLZR23,17WWWmotif,colorcodingVLDB}. Biclique counting is a prominent problem in specific subgraph counting, where general counting algorithms are often not as efficient as specific methods. %Parallelism also plays an important role in subgraph counting, with several algorithms utilizing parallel architectures to improve performance \cite{DBLP:journals/pacmmod/YeLSG24, triangleCPUparallel, sigmodGPUTriangle, DBLP:conf/cluster/RibeiroSL10}. Similarly, parallel biclique counting has been explored \cite{DBLP:conf/icde/QiuLKCG24, BClistYangPZ21}. 
When \(p = 2\) and \(q = 2\), the \((p, q)\)-biclique is referred to as a butterfly. As a special case of general biclique counting, efficient butterfly counting plays a critical role in the development of advanced algorithms that can handle larger graphs and subgraph patterns. Various algorithms have been introduced to efficiently count butterflies in large bipartite graphs, taking advantage of the bipartite structure to reduce computational complexity \cite{LQ19Butterfly, kdd18butterfly}. Recent research has focused on optimizing both memory and time complexity by employing parallel computing techniques \cite{butterflyGPU22, DBLP:journals/vldb/XiaZXZYLDDHM24, DBLP:journals/vldb/WangLLSTZ24}, as well as I/O-efficient methods \cite{DBLP:journals/pacmmod/WangLLS0023} and approximation strategies \cite{tkdd22Butterfly}. Other efforts have been made to extend butterfly counting to uncertain graphs \cite{VLDP21UncertainButterfly, DBLP:journals/vldb/ZhouWC23} and temporal graphs \cite{DBLP:journals/pvldb/CaiKWCZLG23}.

\stitle{$k$-Clique Counting.} Our work is closely related to the \(k\)-clique counting problem in traditional graphs. The first exact algorithm for \(k\)-clique counting, introduced by Chiba and Nishizeki \cite{CN}, relies on a backtracking enumeration method. More recent advances have refined this technique using ordering-based optimizations, such as degeneracy ordering by Danisch et al. \cite{kClist}, and color ordering by Li et al. \cite{LiVldb}. These algorithms efficiently list \(k\)-cliques for small values of \(k\), but their performance degrades as \(k\) increases due to the exhaustive enumeration process. To address the inefficiencies for larger \(k\), Jain and Seshadhri \cite{PIVOTER} proposed \pivot, an algorithm that leverages pivoting techniques for maximal clique enumeration \cite{BK,tomitaTime}. Instead of enumerating all possible \(k\)-cliques, \pivot counts them combinatorially, offering significant performance gains over enumeration-based methods. However, the algorithm may still struggle when applied to large, dense graphs \cite{PIVOTER}. To address this, several sampling-based methods have been developed to improve scalability, including the TuranShadow algorithm and its optimized version \cite{TuranShadow,PEANUTS}, as well as color-based sampling techniques by Ye et al. \cite{ccpath,ye2023efficient}. Despite these efforts, sampling methods tend to become inefficient for larger values of \(k\). Our approach fundamentally  differs from these \(k\)-clique counting studies as we focus on counting bicliques in bipartite graphs, a problem that cannot be directly tackled by the existing \(k\)-clique counting techniques. 

% \stitle{Butterfly Counting.}  
%When \(p = 2\) and \(q = 2\), the \((p, q)\)-biclique is referred to as a butterfly. As a special case of general biclique counting, efficient butterfly counting plays a critical role in the development of advanced algorithms that can handle larger graphs and subgraph patterns. Various algorithms have been introduced to efficiently count butterflies in large bipartite graphs, taking advantage of the bipartite structure to reduce computational complexity \cite{LQ19Butterfly, kdd18butterfly}. Recent research has focused on optimizing both memory and time complexity by employing parallel computing techniques \cite{butterflyGPU22, DBLP:journals/vldb/XiaZXZYLDDHM24, DBLP:journals/vldb/WangLLSTZ24}, as well as I/O-efficient methods \cite{DBLP:journals/pacmmod/WangLLS0023} and approximation strategies \cite{tkdd22Butterfly}. Other efforts have been made to extend butterfly counting to uncertain graphs \cite{VLDP21UncertainButterfly, DBLP:journals/vldb/ZhouWC23} and temporal graphs \cite{DBLP:journals/pvldb/CaiKWCZLG23}.

\section{conclusion}
In this paper, we propose \npivot, a novel and general framework for biclique counting. The \npivot framework not only unifies existing approaches like \bc and \edgepivot but also extends their capabilities by supporting both local and range counting. Our implementation of \npivot leverages a novel minimum non-neighbor candidate partition strategy, which significantly improves the worst-case time complexity compared to previous methods. Furthermore, the inclusion of a novel cost estimator enables the framework to adaptively switch between node-split and edge-split strategies, further enhancing its performance and scalability across different graph structures. Extensive experiments conducted on real-world datasets have demonstrated the efficiency and versatility of \npivot, achieving up to two orders of magnitude improvements over state-of-the-art algorithms.

%In summary, \npivot represents a significant advancement in the field of biclique counting, addressing the shortcomings of existing methods while offering a flexible and scalable solution for various real-world applications.

\bibliographystyle{ACM-Reference-Format}
\bibliography{biclique}

%\newpage
%\input{appendix}
\end{document}